\documentclass[12pt,preprint]{aastex}

\usepackage{emulateapj5}

\slugcomment{Accepted by ApJS}

\shortauthors{Becker, White, et al.}
\shorttitle{The FIRST Bright Quasar Survey South}

\begin{document}

\title{The {\sl FIRST} Bright Quasar Survey III. The South Galactic Cap\altaffilmark{1}}

\author{
Robert~H.~Becker\altaffilmark{2,3,4},
Richard~L.~White\altaffilmark{4,5},
Michael~D.~Gregg\altaffilmark{2,3},
Sally~A.~Laurent-Muehleisen\altaffilmark{2,3,4},
Michael~S.~Brotherton\altaffilmark{3,4},
Chris~D.~Impey\altaffilmark{6},
Frederic~H.~Chaffee\altaffilmark{7},
Gordon~T.~Richards\altaffilmark{8},
David~J.~Helfand\altaffilmark{4,9,10},
Mark Lacy\altaffilmark{2,3},
Frederic~Courbin\altaffilmark{11}
\&
Deanne~D.~Proctor\altaffilmark{3}}
\email{bob@igpp.ucllnl.org}

\altaffiltext{1}{Based on observations obtained with the W.~M.~Keck
Observatory, which is jointly operated by the California Institute of
Technology and the University of California, and the Multiple Mirror
Telescope Observatory, which is jointly operated by the University of
Arizona and the Smithsonian Institution.}
\altaffiltext{2}{Physics Dept., University of California--Davis}
\altaffiltext{3}{IGPP/Lawrence Livermore National Laboratory}
\altaffiltext{4}{Visiting Astronomer, Kitt Peak National
Observatory, National Optical Astronomy Observatory}
\altaffiltext{5}{Space Telescope Science Institute,
3700 San Martin Dr., Baltimore, MD 21218, rlw@stsci.edu}
\altaffiltext{6}{Steward Observatory, U. Arizona}
\altaffiltext{7}{W. M. Keck Observatory}
\altaffiltext{8}{Astronomy \& Astrophysics Center, U. Chicago}
\altaffiltext{9}{Astronomy Dept., Columbia University}
\altaffiltext{10}{Institute of Astronomy, Cambridge University}
\altaffiltext{11}{Pontificia Universidad Catolica de Chile}

\begin{abstract}

We present the results of an extension of the FIRST Bright Quasar Survey (FBQS)
to the South Galactic cap, and to a fainter optical magnitude limit. Radio
source counterparts with SERC $R$ magnitudes brighter than 18.9 which meet
the other FBQS criteria are included. We supplement this list with a modest number of
additional objects to test our completeness for quasars with extended
radio morphologies. The survey covers 589 deg$^2$ in two equatorial strips
in the southern cap. We have obtained spectra for 86\% of the 522 candidates,
and find 321 radio-selected quasars of which 264 are reported here for the
first time. A comparison of this fainter sample with the FBQS sample shows the
two to be generally similar. 

Fourteen new broad absorption line (BAL) quasars
are included in this sample. When combined with the previously identified BAL quasars in our
earlier papers, we can discern a break in the frequency of BAL quasars with radio
loudness, namely that the relative number of high-ionization BAL quasars drops by a factor of four
for quasars with a radio-loudness parameter $R^*>100$.

\end{abstract}


\section{Introduction}
\label{sectionintro}

For the past five years, we have been engaged in an effort to identify the
quasar content of the VLA FIRST survey. Our FIRST
Bright Quasar Survey (FBQS) is motivated both by the high
sensitivity and excellent
astrometry provided by FIRST. To date, there have been two releases of
quasars discovered by the FBQS (Gregg et al.\ 1996, hereafter
\citeauthor{gregg96}; White et al.\ 2000, hereafter \citeauthor{white00}).  Both
have presented quasars found in the north Galactic cap down to a
limiting E ($\sim R$) magnitude of 17.8. In this paper, we are
presenting the extension of the survey to the south Galactic cap.
Although the survey area in the south is considerably smaller (589
square degrees vs the 2682 square degrees covered in \citeauthor{white00}),
we have extended the optical magnitude limit to
$R=18.9$.

Except for the fainter magnitude limit, the selection criteria used to
generate the list of 522 candidates are largely the same as those
used in \citeauthor{white00}, but with two additions designed to test our
completeness to quasars with extended radio morphologies.
With spectra now in hand for
86\% of the candidates, we have identified 321 radio-selected quasars,
of which 264 are newly discovered.
In this work we present the vital statistics of the quasars and
nonquasars alike and contrast this sample with the brighter quasars
found previously by the FBQS.

\begin{figure*}
\plotone{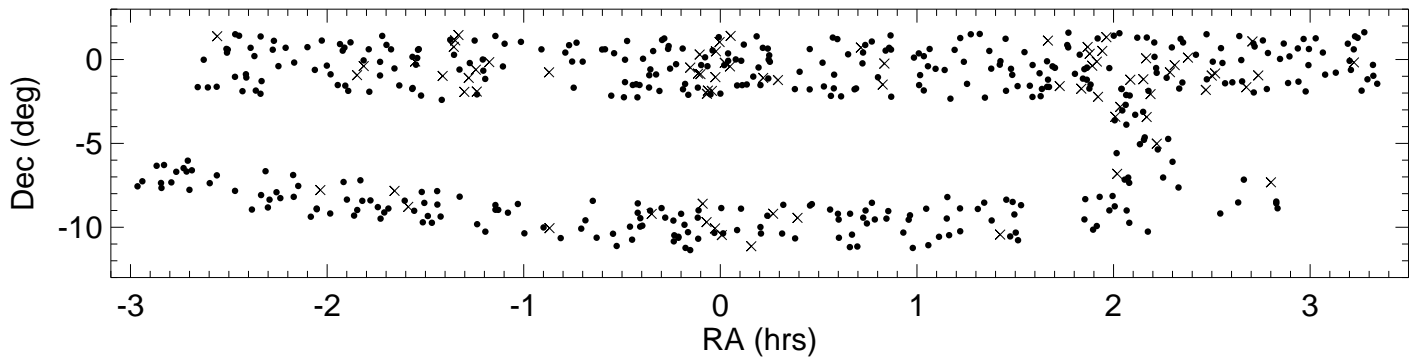}
\caption{
Distribution of the 522 FBQS candidates on the sky.  Dots are
the 447 objects with spectral identifications; x's mark the 75 objects without
spectra. The gaps centered at $01^{\rm h}45^{\rm m}$ and (less obviously)
at $02^{\rm h}20^{\rm m}$ result from missing SERC plate scans.
\label{figskydist}
}
\end{figure*}

\section{The Sample}
\label{sectionsample}

As in the previous renditions of the FBQS, the FIRST survey serves as
the primary catalog (\citealp{becker95,white97}; see also the
FIRST home page at \url{http://sundog.stsci.edu}).  In the south
Galactic cap, the sky coverage is limited to two strips of sky, one
along the equator and a second some 7 to 10 degrees south of the
equator. Both of these areas coincide with regions slated for coverage
by the Sloan Digital Sky Survey \citep{gunn92}. In addition, a short
segment was added at RA $\sim2$~hrs to connect the two disjoint SDSS areas.
Candidate quasars were
identified by comparing the FIRST survey catalog to the APM catalog of
optical sources on the UK Schmidt survey of the southern sky
\citep{mcmahon92}. The spatial distribution of the candidates is shown
in Figure~\ref{figskydist}.  There are a few gaps in the coverage created
where the corresponding UK Schmidt plates were never scanned by APM.
The missing plates are
f684 (centered at $\alpha=1^{\rm h}42^{\rm m}$, $\delta=-10^\circ$),
f686 ($2^{\rm h}22^{\rm m}$, $-10^\circ$),
f688 ($3^{\rm h}02^{\rm m}$, $-10^\circ$), and
f760 ($3^{\rm h}02^{\rm m}$, $-5^\circ$).

We followed many of the procedures described in detail in
\citeauthor{white00}.  Perhaps most importantly, we used the FIRST
catalog positions to calculate a new astrometric solution for the UK
Schmidt plates. In \citeauthor{white00}, we applied a plate-by-plate
calibration derived from the APS catalog \citep{pennington93} to the
APM photometry.  The APS data is limited in the southern Galactic cap,
so for this paper we applied a similar photometric calibration using
instead an early version, 2.1.1, of the Guide Star
Catalog-II.\footnote{Information on the GSC-II catalog is available at
\url{http://www-gsss.stsci.edu/gsc/compass/GSC2export/GSC2Export.html}.}
The photometric calibration used stars in the vicinity of FIRST sources
rather than the FIRST optical counterparts themselves. This is important
because the FBQS quasars are variable, and Malmquist bias between
the APM and GSC-II epochs causes the GSC-II magnitudes for quasars to
appear systematically fainter than the APM magnitudes \citep{helfandvar}.
A comparison between the GSC-II and APS magnitude calibrations in the
north Galactic cap indicates excellent agreement, and we estimate that
the calibrated magnitudes in this paper have an rms accuracy of
$\sim0.15^{\hbox{m}}$ on both the $R$ and $B$ plates, which is similar
to the uncertainties for the $E$ and $O$ magnitudes in \citeauthor{white00}.

Any optical counterpart to a FIRST radio source which met the following
conditions was considered a quasar candidate:
\begin{itemize}
\item The radio and optical positions must coincide to better than 1.2
arcsec.
\item The optical counterpart must be classified as stellar on at least
one of the two plates.
\item The recalibrated, extinction-corrected optical SERC $R$ magnitude
(5900--6900~\AA) on the
UK Schmidt red plate must be brighter than 18.9.
\item The UK Schmidt color must be bluer than $B-R = 2$.
\end{itemize}
These criteria resulted in a candidate list of 505 optical counterparts.
In addition, we included another seventeen candidates which failed the above
criteria but represented one of the two situations described below.

If a FIRST radio
source was elongated such that its radio centroid differed from the
radio peak by more than 1~arcsec, the peak position was calculated and
used in a second search for an optical counterpart.  This resulted in
the selection of another 5~candidates, 3~of which turned out to be
quasars. These candidates are identified with the comment `CJ'
(core-jet) in the tables below.

In addition, for all radio doubles separated by $<30\arcsec$ in the survey
area, optical counterparts were identified along the line joining the
two radio sources under the assumption that the radio double has an
FR~II radio morphology in which the radio flux density from the core
falls below the FIRST survey limit --- i.e., radio-loud quasars with
extremely small core-to-lobe ratios. The positional criteria were
determined from an analysis of $\sim8000$ optical counterparts to
double radio sources in the FIRST survey \citep{mcmahon01}; we include
sources within $\pm1.5\arcsec$ of the line between the radio components
and within $\pm2\arcsec$ of the flux-weighted midpoint along the connecting
line.  This search added another 12 candidates to the program, of which
10 were found to be quasars. These candidates are designated with the
comment `DBL' (double) in the lists given below.

A simple search for quasars at the location of the core emission finds
96\% of the radio-selected quasars. A more diligent search as outlined
above finds an additional 4\% of the quasars, and the majority of these
are faint: for our limiting magnitude of $R=17.8$ in the northern
survey,
the results here imply that we are missing less than one quasar per 100 deg$^2$
as a consequence of its morphology, or $\sim3$\% of the total bright quasar
population. It is worth noting that
the efficiency of finding quasars with the expanded procedure (57\% of
the candidates turn out to be quasars) is comparable to the efficiency
of searching the core positions (68\% of which are quasars).

\begin{figure*}
\plottwo{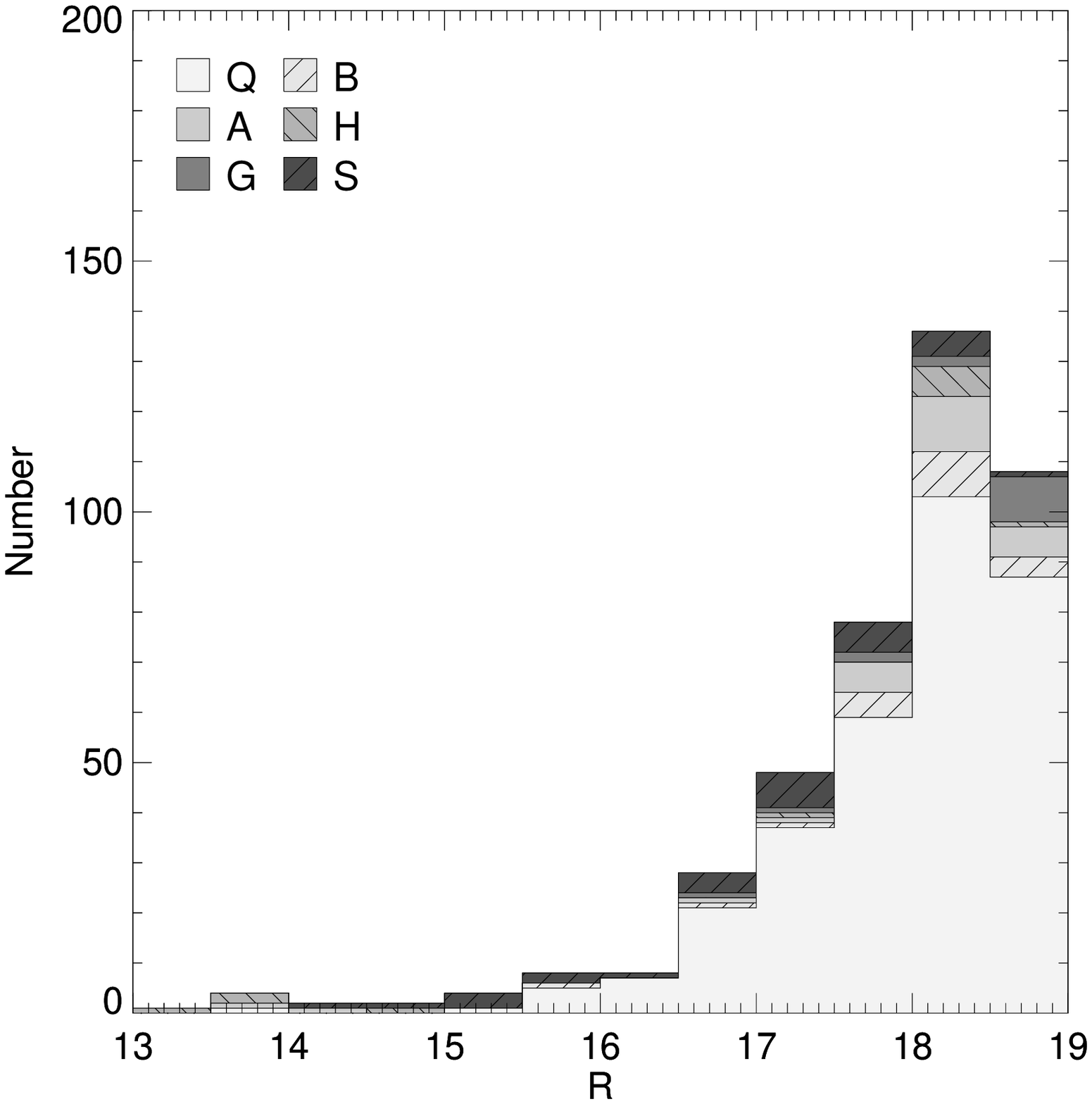}{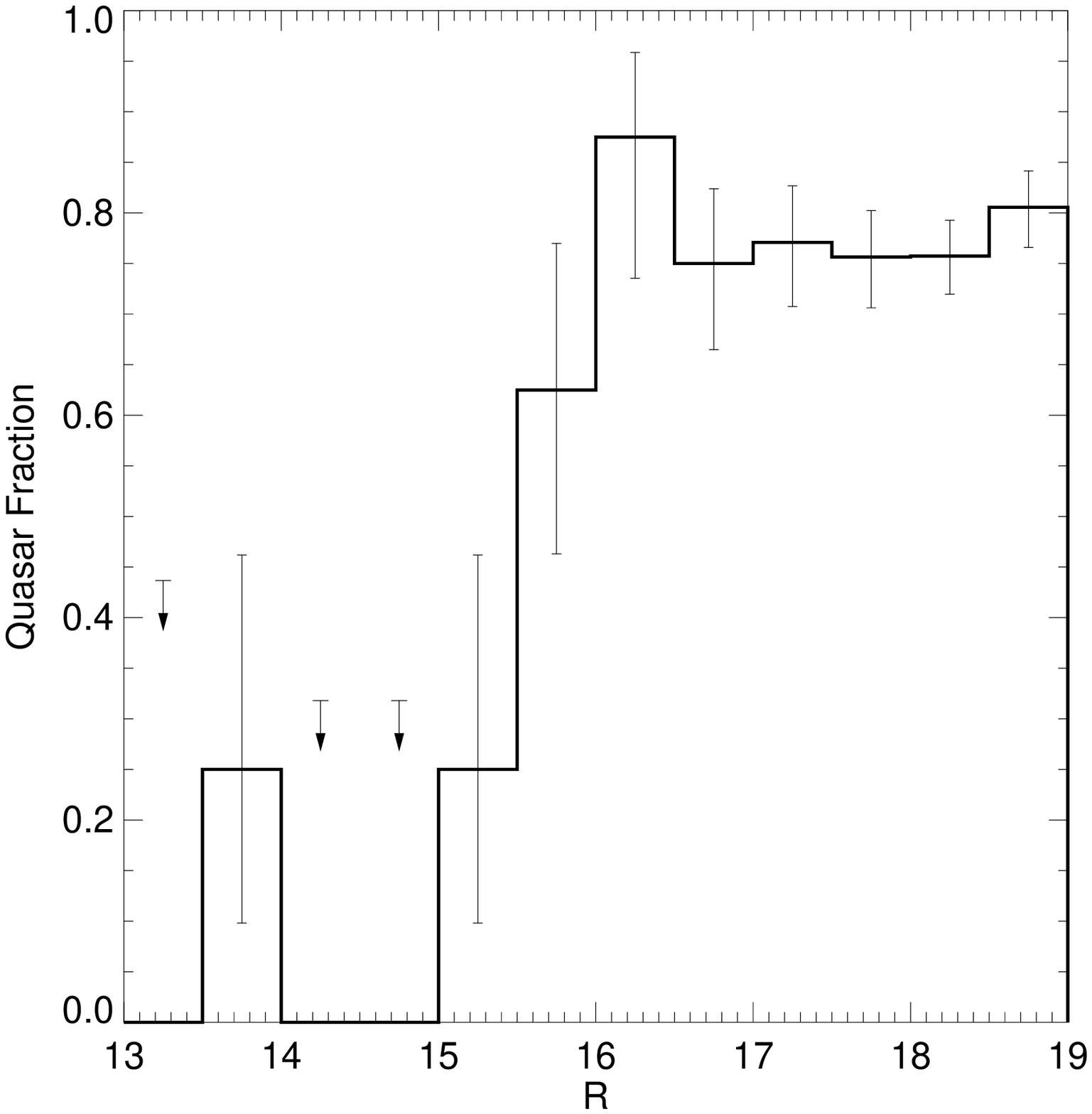}
\caption{
(a) A histogram of extinction-corrected SERC $R$ magnitudes for
FBQS candidates with identifications: Q=quasar, B=BL Lac object,
A=narrow-line AGN, H=star-forming H~II region galaxy, G=normal
galaxy (no emission lines), and S=star (see \S4 for details.)
(b) The fraction of quasar identifications as function
of $R$ magnitude.
\label{figmaghist}
}
\end{figure*}

\begin{figure*}
\plottwo{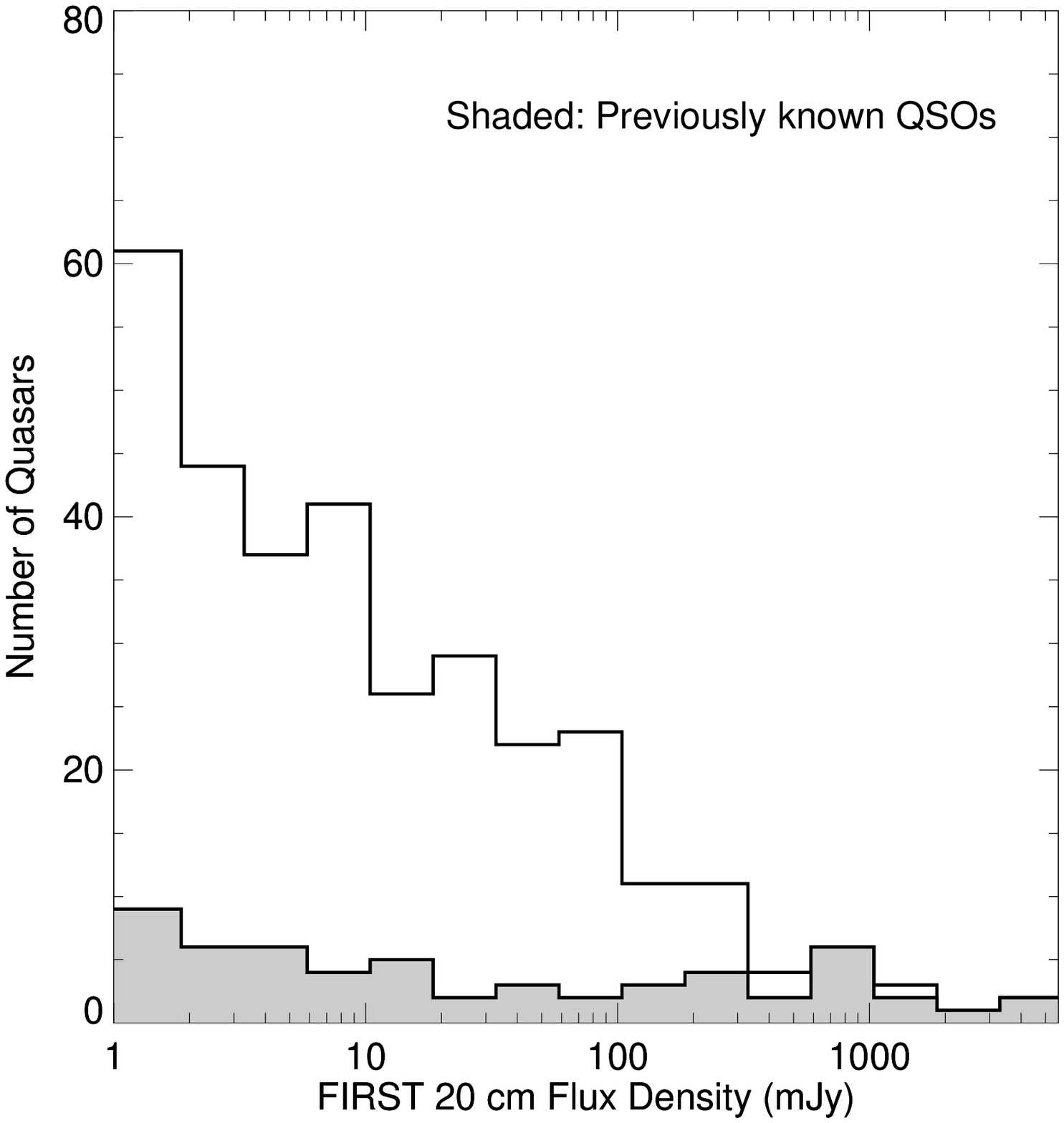}{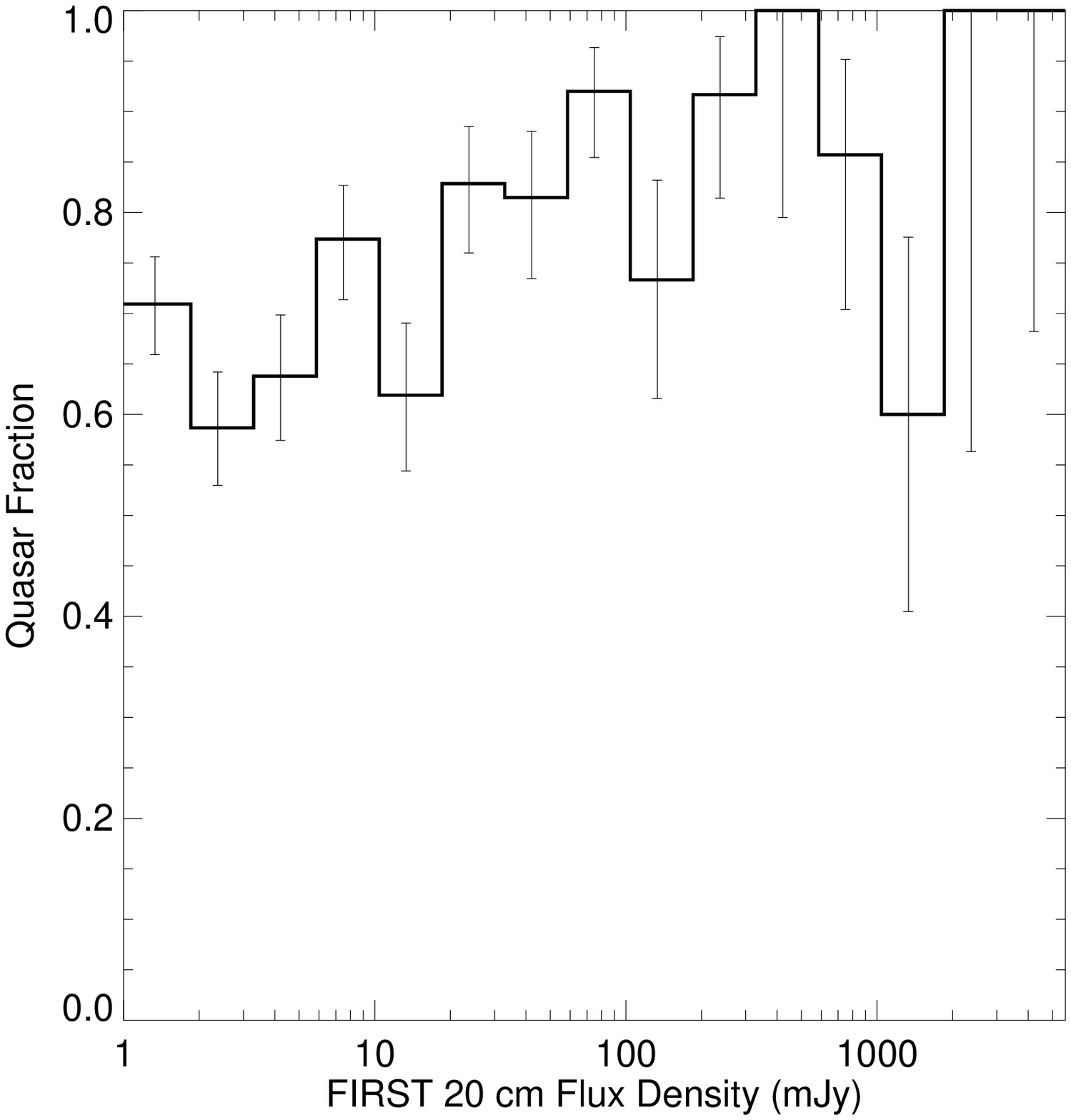}
\caption{
(a) A histogram of FIRST integrated radio flux densities
of FBQS quasars.  Shaded: previously known quasars.
(b) The fraction of FBQS candidates identified as quasars as function
of radio flux density. The flux densities come from the FIRST
catalog and so include only the core radio emission.
\label{figfluxhist}
}
\end{figure*}

\begin{figure*}
\plottwo{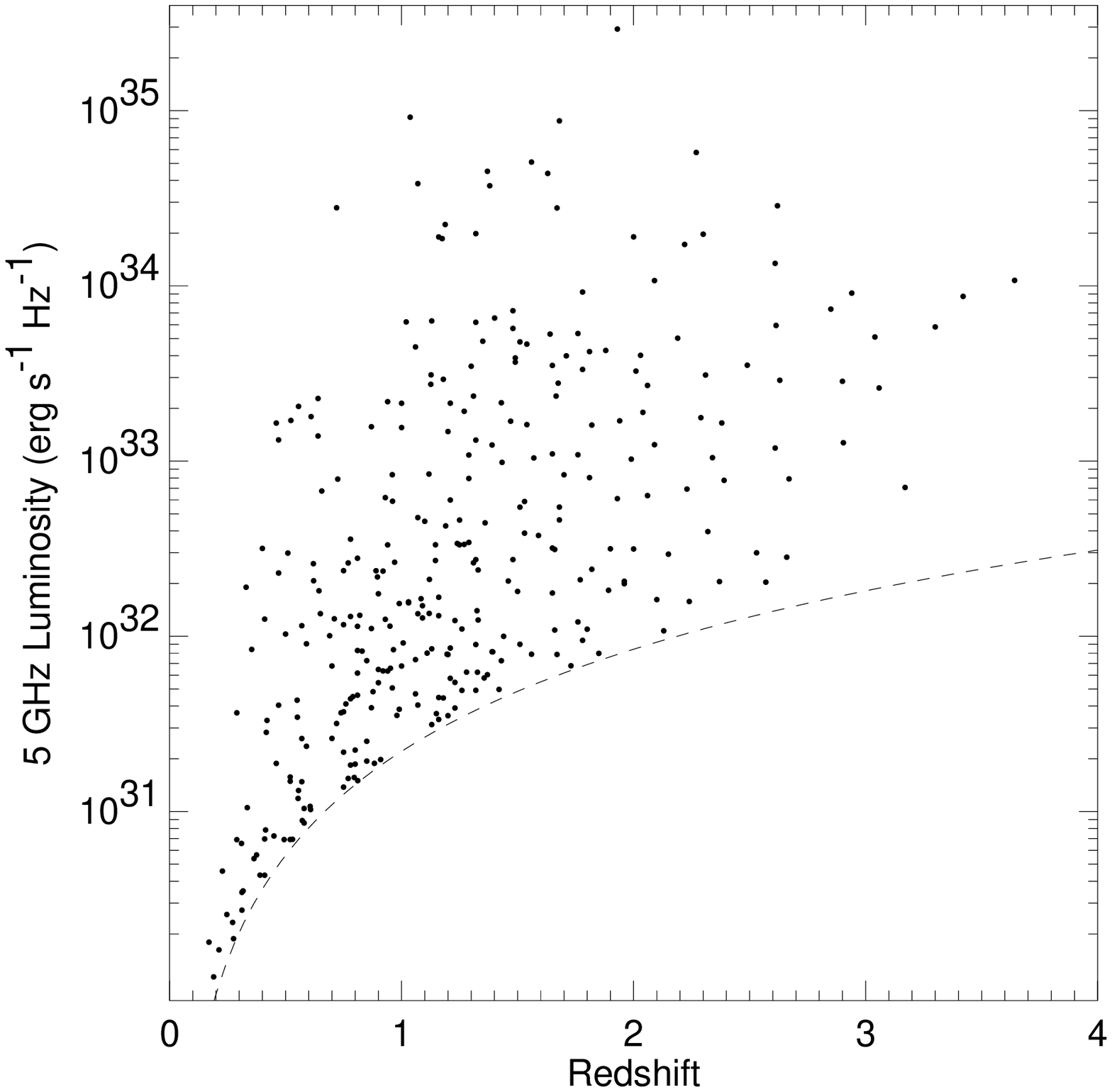}{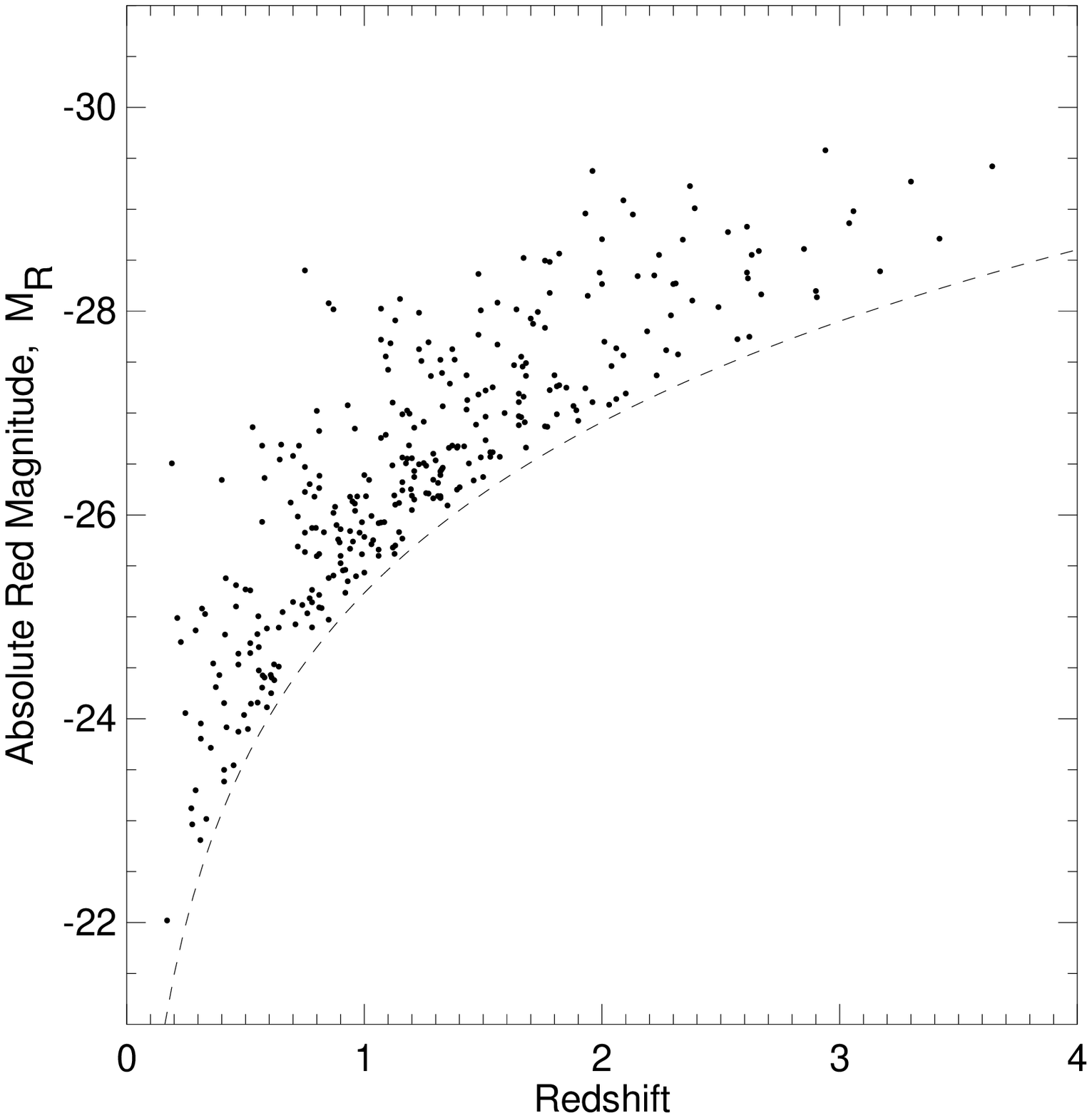}
\caption{
FBQS quasar luminosities versus redshift in the radio and optical.
(a) Radio luminosity at a rest frequency of 5~GHz versus redshift,
using spectral index $\alpha=-0.5$.
(b) Absolute red magnitude versus redshift, using spectral index
$\alpha_{opt}=-1$ for the $K$-correction.  Dashed lines show the 1~mJy
FIRST detection limit and the $R=18.9$ APM magnitude limit.  The radio
luminosities have a much larger dynamic range and do not crowd as
closely against the detection limit as do the optical magnitudes.
\label{figlumz}
}
\end{figure*}

\subsection{Sample Biases}

In \citeauthor{white00}, we used a series of histograms and plots to
illustrate the degree to which our selection criteria might be biasing
the sample. For example, Figures~2a and~2b in \citeauthor{white00} were
histograms displaying the number of quasars as a function of angular
separation between the radio and optical positions and the fraction of
candidates which turned out to be quasars, also as a function of angular
separation. The same histograms for this new sample are effectively
identical; i.e., they show that very few quasars are found near the
limiting separation of 1.2 arcsec.  From this we still conclude that few
quasars are lost due to a positional mismatch. The one obvious exception,
quasars with radio jets, have now been searched for (as described above)
with the result that only three additional quasars were found. While our
method for going after core-jet sources is not foolproof --- for example,
we will still miss quasars where the jet is brighter than the core ---
it does suggest that these objects constitute a very small subsample
of quasars.

The distribution of $R$ magnitudes for the current sample is quite different
from that of \citeauthor{white00} for the simple reason that we have chosen to
go $\sim1.1$ magnitudes fainter in the SGP. Hence the histogram of the
number of quasars vs.\ $R$ magnitude in Figure~\ref{figmaghist}a rises sharply down to
the survey limit of 19th magnitude. Figure~\ref{figmaghist}b shows the success rate for
finding quasars as a function of $R$. As in \citeauthor{white00}, the fraction
increases with increasing magnitude before leveling off at 70\% for
candidates fainter than $R$ of 16.5. The lower success rate at brighter
magnitudes is largely due to optically bright galaxies which the APM
catalog has misclassified as stellar.

Going deeper has not resulted in a significant difference in the color
of the quasars found. As shown in Figure~5a from \citeauthor{white00}, most of
the quasars in the previous sample have O-E between 0 and 1. The same
range is applicable for the $B-R$ colors of the current sample; of candidates
with $B-R > 1.5$, only 10\% turn out to be quasars. There
appears to be a shift of 0.1 magnitudes in the peak of the distribution
(the current sample is slightly bluer), but this is possibly a consequence
of the different bandpasses. Likewise a color-redshift plot for the
current sample is very similar to the equivalent plot in Figure~6 of
\citeauthor{white00}.

In Figure~\ref{figfluxhist}a we show a histogram of the number of quasars vs.\ the
20~cm radio flux density.  As in \citeauthor{white00}, the number of quasars
rises down to the flux density limit of the FIRST survey.
Figure~\ref{figfluxhist}b
demonstrates that we are nearly as efficient at finding radio-weak quasars
as radio-bright quasars. In Figures~\ref{figlumz}a
and~\ref{figlumz}b we show plots of radio luminosity and absolute $R$ magnitude vs.\
redshift. The radio luminosity plot is virtually indistinguishable from
Figure~8a in \citeauthor{white00}, an unsurprising result given that both samples have the
same radio flux density limit.
The plot of $M_R$ (Fig.~\ref{figlumz}b) is strikingly
different since here we are going 1.1 magnitudes deeper over a much
smaller area.  In this sample there is a relative paucity of very luminous
quasars and a much stronger representation of lower luminosity quasars.

\section{Optical Spectroscopy}
\label{sectionspectra}

Spectra were collected at a variety of observatories, including the
Lick Shane 3-m telescope, the `classic' MMT ($6\times1.8$-m), the 10-m
Keck-II Observatory, and the ESO 3.6-m telescope. Instrument parameters
for the first three observatories are given in Table 1 of \citeauthor{white00}; for
the ESO 3.6-m, we used the EFOSC2 (ESO Faint Object Spectrograph and Camera).
The observations
were made without regard to atmospheric transparency or seeing, so
the flux calibration of the resulting spectra are not absolute.  Where
possible, the slit was oriented at the parallactic angle to minimize
differential slit losses.  Since the data were taken at a number of
different sites with dissimilar spectrographs, the resolution and
wavelength coverage are not uniform. In all cases the spectra were
reduced using standard IRAF routines.

\section{Spectroscopic Results}
\label{sectionresults}

Spectra were obtained (or identifications determined from the literature)
for 447 of the 522 quasar candidates, and
were classified using the same conventions described in detail in
\citeauthor{white00}. Briefly, we classified spectra as quasars if there were
any broad emission lines with rest-frame equivalent widths greater
than 5~\AA; objects with weaker broad lines were classified as BL Lacs.
We classified a spectrum as a narrow-line AGN if narrow lines were present
in the spectra with ratios consistent with those expected for an AGN
(i.e., [O~III]$\lambda5007$/H$\beta$ greater than 3 or
[N~II]$\lambda6583$/H$\alpha$ greater than
0.6). Any spectra with narrow lines failing to meet these criteria were
classified as H~II/star-forming galaxies. Spectra with very weak or no
emission lines were classified as normal galaxies, unless the Ca H~\&~K
4000~\AA\ break was less than 25\% in which case they
were assigned a classification of BL Lac.
Spectra at zero redshift were deemed stars. In all, the distribution of
objects was 321 quasars, 23 BL Lac, 31 AGN, 23 starburst, 17 normal
galaxies, and 32 stars. The latter are roughly consistent with the expected
chance coincidence rate (although a few are real detections of stellar radio
sources -- see \citealp{helfandstars}); note that the putative double radio sources
are overrepresented by a factor of $\sim3$ in the stellar list, consistent with
the factor of $\sim3$ increase in the accepted matching area (12~arcsec$^2$ vs. 4.5~arcsec$^2$).

The fractions of the sample that are narrow-line AGN, BL Lacs, and normal
galaxies are all roughly consistent with the fractions derived in our sample
limited to $R\le17.8$. The fraction of H~II galaxies, however, is lower by
a factor of 3.5, a result of the fainter optical magnitude limit of this
sample. The radio flux densities of most of the star-forming galaxies are very
close to the FIRST threshold (85\% have $S_p<3.0$~mJy). For a constant
radio-to-optical flux ratio, lowering the optical magnitude limit by 1.2
magnitudes would render most of these objects undetectable. Finally, the
fraction of quasars is considerably higher in this fainter sample, reflecting
their steep $N(m)$ distribution.

In Tables 1--6 we list all the candidates with spectroscopic
classifications. For each object we include the FIRST catalog RA and Dec
(J2000), the recalibrated and extinction-corrected $R$ and $B$ magnitudes,
the red extinction correction $A(R)$, and the FIRST peak and integrated
radio flux densities. The radio-optical positional separation and APM
star-galaxy classification (used to define the sample) are also given.
The objects have been segregated into 6 tables by their optical
spectral classification.  Quasars are listed in Table~1, narrow line
AGN in Table~2, BL Lac objects in Table~3, H~II/star-forming galaxies
in Table~4, galaxies without strong emission lines in Table~5, and
stars in Table~6. Table~7 lists the objects for which spectra have not
yet been obtained.

In Tables 1--5, we list the measured redshift (except for the 2/3 of the BL
Lacs for which none was derivable from the spectrum). We use
the redshift to calculate for each object the radio luminosity $L_R$ at
a rest frequency of 5~GHz (assuming a radio spectral index of $-0.5$),
the absolute $B$ magnitude $M_B$, and, as a measure of radio loudness,
the ratio $R^*$ of the 5~GHz radio flux density to the
2500~\AA\ optical flux in the quasar rest frame (assuming an optical
spectral index of $-1$ and the definition given by \citealp{stocke92}). The
cosmological parameters
$H_0=50\,\hbox{km}\,\hbox{s}^{-1}\hbox{Mpc}^{-1}$, $\Omega=1$, and
$\Lambda=0$ were used for the luminosity calculations. There is also a
comment column, which notes the details of particular interest such as
whether objects were selected as core-jet or double sources, the
presence of broad absorption lines (BALs) or damped Lyman alpha absorption
lines, whether the object was previously known or, in NED, associated
with a ROSAT or IRAS source.

The spectra for all the objects identified as quasars are displayed in
the figure at the end of the paper.

\begin{figure*}
\plottwo{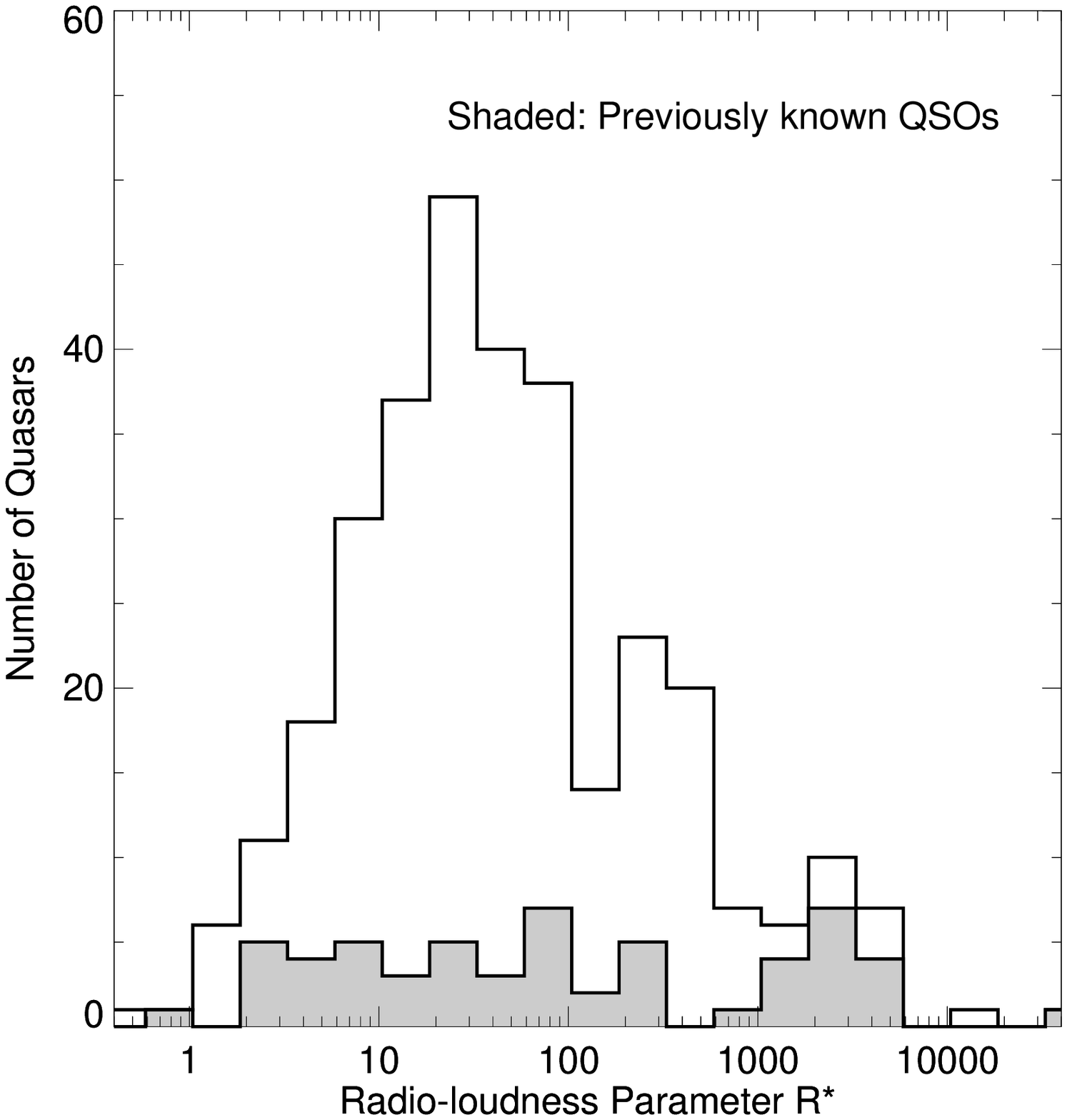}{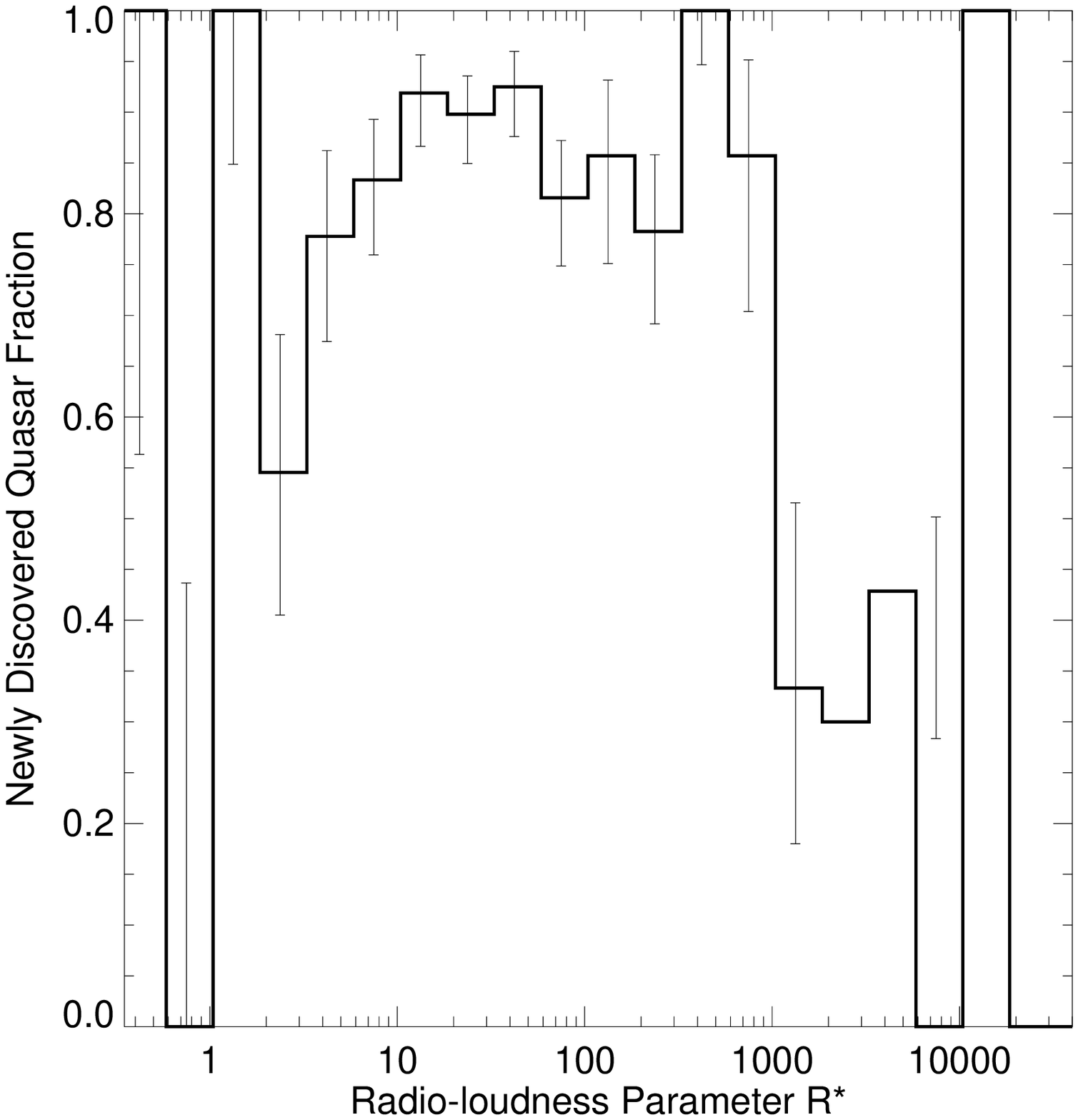}
\caption{
(a) A histogram of radio-optical ratio $R^*$ \citep{stocke92} for FBQS
quasars.  Shaded: previously known quasars.
(b) The fraction of quasars that were newly discovered
versus $R^*$.  The FBQS is increasing the number of known objects in
the radio-quiet/radio-loud transition region ($R^*=1$--100) by a large
factor.
\label{figrhist}
}
\end{figure*}

\begin{figure*}
\epsscale{0.45}
\plotone{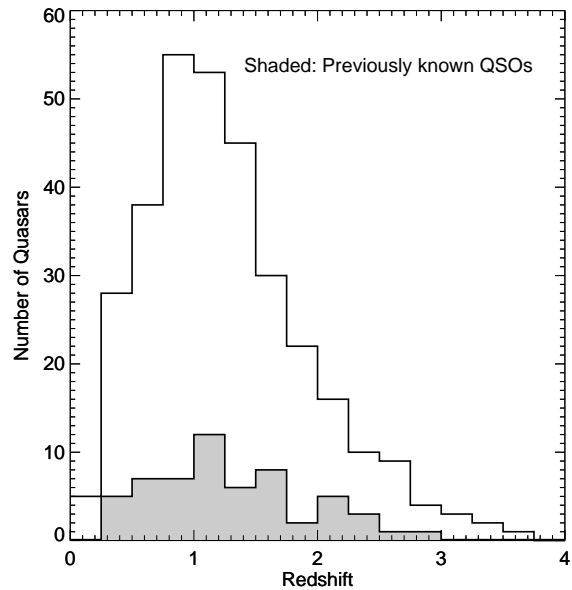}
\epsscale{1}
\caption{
Histogram of redshifts for FBQS quasars.  Shaded: previously known quasars.
\label{figzhist}
}
\end{figure*}

\begin{figure*}
\plottwo{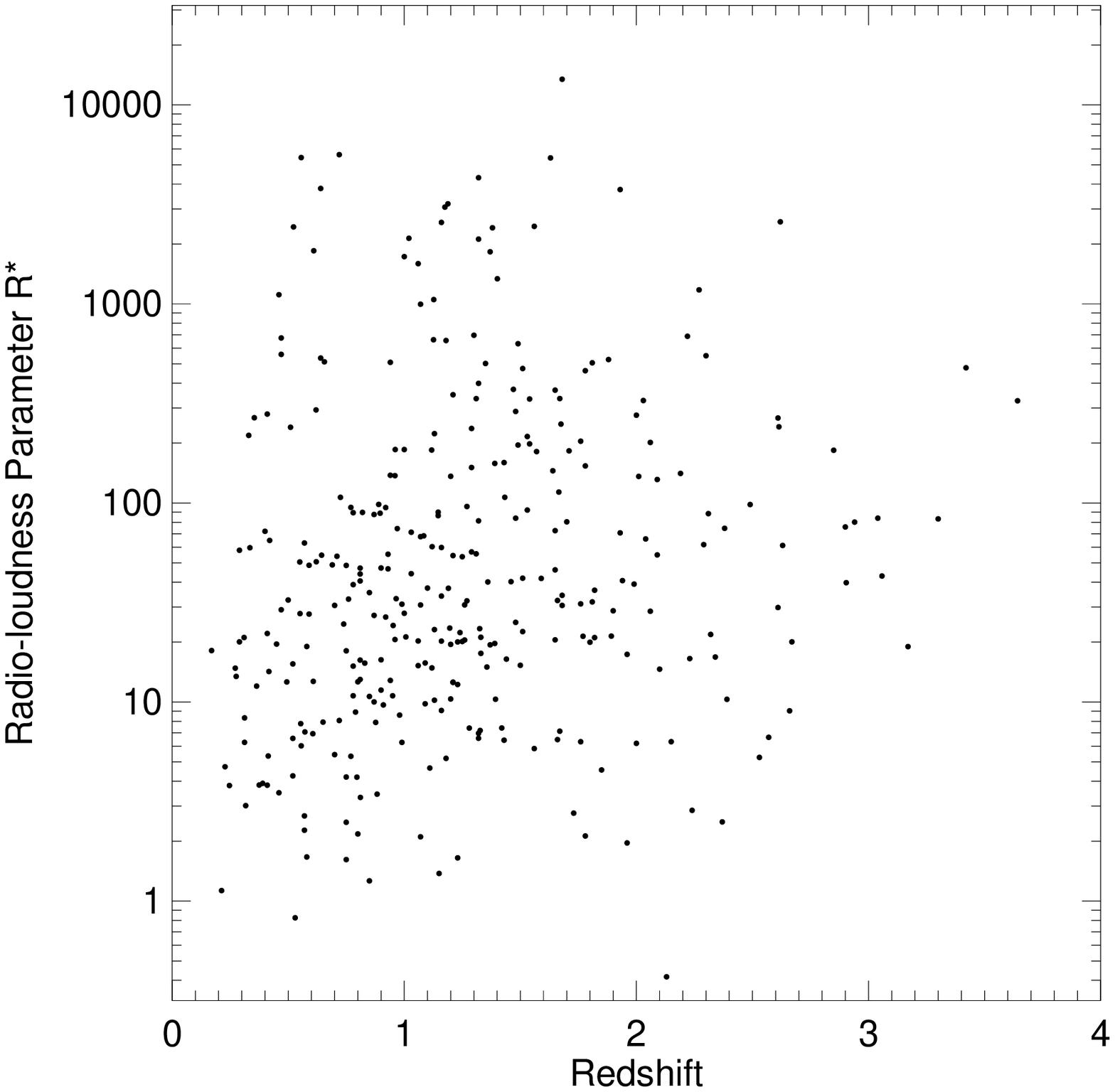}{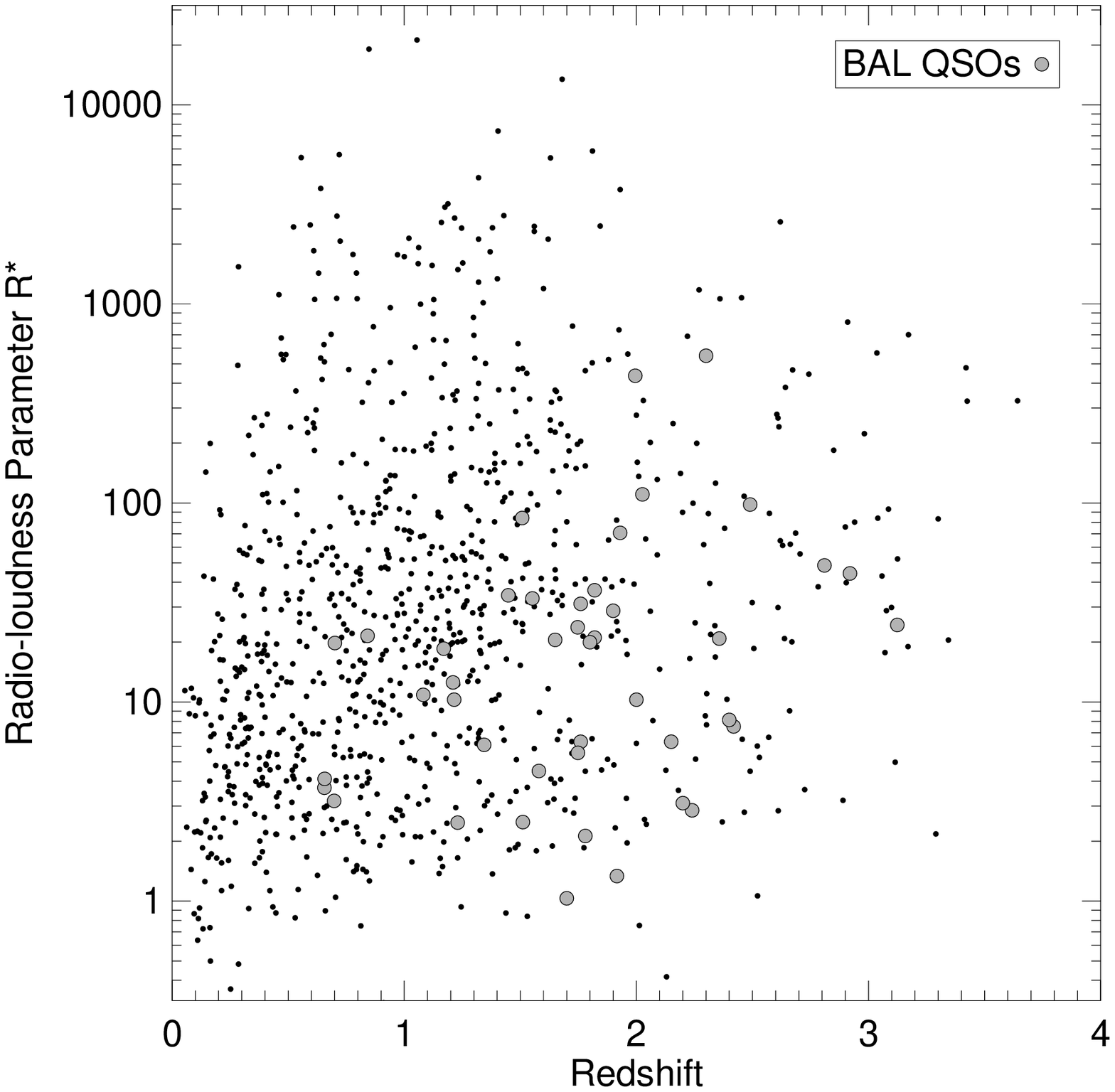}
\caption{
Radio-optical ratio $R^*$ (rest-frame ratio of the 5~GHz radio flux
density to the 2500~\AA\ optical flux; \citet{stocke92}) versus redshift
for FBQS quasars.  (a) Quasars from this paper.  (b) Quasars from this
paper combined with those from Paper~II.  Broad-absorption line quasars
are indicated; they are evidently more common among quasars with $R^*<100$
than among the radio-louder objects.
\label{figrz}
}
\end{figure*}

\begin{figure*}
\plotone{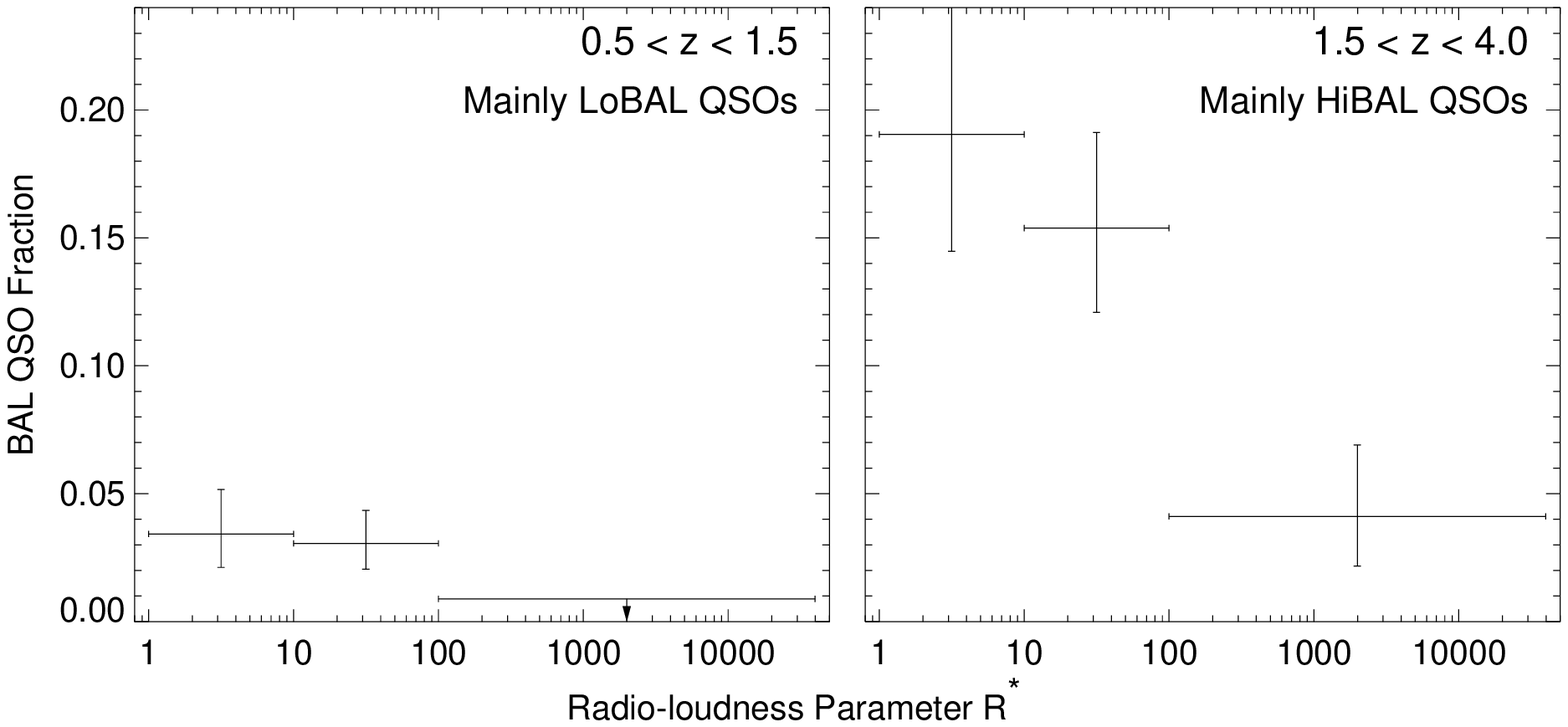}
\caption{
The fraction of BAL quasars as a function of radio-loudness
for the combined samples from this paper and Paper~II.  The left panel
shows the distribution of 12 BAL QSOs (11 LoBALs, 1 HiBAL) among 507
quasars with $0.5 < z < 1.5$.  The right panel the distribution of 31
BAL QSOs (6 LoBALs, 25 HiBALs) among the 245 quasars with $z > 1.5$.
The error bars show the uncertainty in the BAL fraction computed from
the binomial distribution.
Low BALnicity objects classified as `BAL?' are included;
if they were excluded, the plotted points for the $R^*=$10--100 bins
and the $R^*>100$ bin in the right panel would shift
downward by $\sim1/2\sigma$.
The BAL quasar $R^*$ distribution differs from
that for the non-BAL quasars at greater than 99\% confidence for the
$z>1.5$ sample.
\label{figbalfrac}
}
\end{figure*}

\section{Discussion}
\label{sectiondiscussion}

This new quasar sample contains half as many quasars as presented in
\citeauthor{white00}.  This is in part a reflection of two competing
factors; a reduced survey area (600 square degrees vs.\ 2700 square
degrees) and a fainter magnitude limit ($R=18.9$ vs.\ $E=17.8$). If we
restrict the comparison to quasars brighter than $R=17.8$  and
normalize for the area surveyed, we still find a unexpectedly small
number of quasars in the new sample.  For example, the areal densities
(quasars per deg$^{2}$) of quasars with $R < 17.8$ and $0.3 < z < 2.2$
in \citeauthor{white00} and this paper are 0.203 and 0.149
respectively, i.e., this paper has 27\% fewer bright quasars than
expected based on \citeauthor{white00}.  This difference is significant
at the 2.9 sigma level.  Since quasars have a very steep
number/magnitude relation, even a small systematic error in the
photometry of the two samples could account for much of the
discrepancy.  In fact, the entire effect could arise if this paper
magnitudes were only 0.15 magnitudes too faint.  Although we don't
believe the systematic errors are that large, even an error of 0.1
magnitudes could reduce the discrepancy to below 2 sigma.

Of course, the radio flux density limits for \citeauthor{white00} and
this paper are nominally the same. Since the radio loudness is defined
in terms of the ratio of radio to optical luminosity, the extra quasars
found by going to fainter optical limits are, on average, more
radio-loud. This can be seen in Figure~\ref{figrhist} which shows a
histogram of the number of quasars as a function of radio loudness.
The histogram peaks at $R^*$ greater than 10, whereas in the previous
sample (see Fig.~15a in \citeauthor{white00}) the equivalent histogram
peaks at 5. Therefore differences between the two samples will, in
part, reveal factors that depend on radio loudness.  Going to fainter
optical magnitudes does not appreciably increase our sensitivity to
nearby quasars because the rapid evolution of quasars means there are
few nearby quasars to discover, and very low-luminosity, nearby quasars
will not look stellar and hence do not get into the survey.
Consequently, it is not surprising that this new sample tends to select
a more distant set of quasars. In Figure~\ref{figzhist} we show the
histogram of the quasar redshifts.  While in \citeauthor{white00} this
distribution was flat between redshifts of 0--1.2 (see Fig.~12 in
\citeauthor{white00}), this sample shows a significant decline for
redshifts below 0.7.

One attribute of quasars which appears to correlate with radio-loudness
is the likelihood of broad absorption lines (BALs) in a quasar
spectrum.  For a long time, it was thought that BALs only occur in
radio-quiet quasars \citep{stocke92}. More recently, quasar searches
based on the FIRST and NVSS radio surveys have found a number of
radio-loud BAL quasars \citep{becker97,brotherton98,becker00}. However,
the size of the samples has limited our ability to quantify how BAL
frequency depends on radio-loudness. With this sample, it is worth
revisiting this question.  In Figure~\ref{figrz}(b) we plot
radio-loudness vs.\ redshift for the combined samples from this work
and \citeauthor{white00}.  The BALs are highlighted in the figure.  It
is clear from the figure that the BALs straddle the
radio-loud/radio-quiet divide but do not seem to occur as often in the
most radio-loud segment of the population.

The classification of quasars as BAL QSOs is somewhat problematic. One
approach is to calculate a `BALnicity' index following the prescription
of \citet{weymann91}, although in many ways this approach is too
conservative \citep{becker00}. Table~8 gives the BALnicity indices for
all our candidate BAL quasars; those with zero BALnicity values (that
we nonetheless believe are likely to be BAL QSOs) are classified as
`BAL?'.  Believing in a democratic approach, we
show all the spectra in the sample in Figure~\ref{figspectra} so the
reader can see for herself the features we are willing to call BALs.
The figure also serves other purposes since many other phenomena are
present in quasar spectra.

Figure~\ref{figbalfrac} shows the fraction of BAL quasars (including
4 objects classified as `BAL?') as a function
of radio-loudness for the combined samples from \citeauthor{white00} and
this paper.  If the `BAL?' objects were excluded,
the plotted points would shift by less than the error bars.
In examining Figures~\ref{figrz}
and~\ref{figbalfrac}, it is worthwhile to note that there are three
distinct types of BAL quasars: high ionization objects (HiBALs) that
show broad absorption from C~IV and Si~IV; low ionization objects
(LoBALs) that, in addition to C~IV and Si~IV, also show absorption due
to Mg~II and occasionally Al~III $\lambda1858$; and a rare class of
LoBALs (FeLoBALs) that show superposed absorption from metastable
excited states of Fe~II. Since C~IV does not move into our spectral
window for $z<1.5$, identified HiBALs are limited to redshifts greater
than 1.5.  So while some of the low redshift quasars may be HiBALs,
they cannot be recognized as such.  Likewise some of the $z>1.5$ HiBALs
might actually be LoBALs or FeLoBALs but since Mg~II is shifted into
the near IR at these redshifts, they cannot be so identified unless
absorption by Al~III is also present.  For the most part, the fraction
of BALs found among $z>1.5$ quasars can be taken as a proxy for the
fraction of HiBALs.  And by the same token, the fraction of quasars
showing BALs in the redshift range of 0.5--1.5 can be taken as a proxy
for the fraction of LoBALs.

In \citet{becker00}, we found the fraction of quasars which are HiBALs
and LoBALs (including FeLoBALs) among radio-selected quasars with
$R<17.8$ to be 18\% and 3\% respectively.  If we combine the current
sample with those from \citeauthor{white00} and consider only quasars
with a radio-loudness parameter $R^* < 100$, we get very similar
numbers (17\% and 3\%; see Fig.~\ref{figbalfrac}). But for quasars with
$R^* > 100$ the fraction of HiBALs and LoBALs drops to 4\% and $<1.5$\%
respectively.  (If the low-BALnicity `HiBAL?' objects are omitted, the
fraction of $R^*>100$ quasars that are HiBALs drops to 3\%.) While the
apparent drop in the fraction of LoBALs among the most radio-loud
quasars is not statistically significant (due to the small number of
quasars that are LoBALs), the decrease in HiBALs is significant.  We
can rule out at 98.9\% confidence the hypothesis that the $R^*<100$ and
$R^*>100$ BAL fractions are the same for the $z>1.5$ sample.  A
two-sided Kolmogorov-Smirnov test \citep{numrecipes} indicates the
probability that the $z>1.5$ BAL and non-BAL quasars have the same
$R^*$ distribution is only 0.0053.

\section{Editorial}

The utility of quasar surveys has evolved with time. Today, with well
over 10,000 quasars in the literature and many more to appear soon,
the rationale for doing surveys is quite different than in the past.
The need for ever larger surveys is being driven by the desire either to
find large numbers of rare quasars or to nail down subtle correlations
between quasar properties with massive numbers. As an example of the
former, with this addition to the FBQS we are just beginning to discern
the dependence of the BAL phenomena on radio-loudness. The fact is that
current samples are still not large enough to provide quantities of very
rare classes of quasars sufficient for reliable statistical analyses.
Other areas of quasar research that are still limited by current sample
size are gravitational lensing, the frequency and nature of damped
Lyman alpha absorbers, the frequency and longevity of binary quasars,
and quasar large-scale structure.  Perhaps in five years when the SDSS
100,000 quasars and the 2dF 30,000 quasars are publically available,
we may finally stop searching for more quasars. But we sincerely doubt it.

\acknowledgments

We acknowledge extensive
use of the NASA/IPAC Extragalactic Database (NED), which is operated by
the Jet Propulsion Laboratory, Caltech, under contract with the
National Aeronautics and Space Administration.  The success of the
FIRST survey is in large measure due to the generous support of a
number of organizations.  In particular, we acknowledge support from
the NRAO, the NSF (grants AST-98-02791 and AST-98-02732), the Institute
of Geophysics and Planetary Physics (operated under the auspices of the
U.S. Department of Energy by Lawrence Livermore National Laboratory
under contract No.~W-7405-Eng-48), the Space Telescope Science
Institute, the National Geographic Society (grant NGS
No.~5393-094), and Columbia University. F.~Courbin
acknowledges financial support from Chilean grant FONDECYT/3990024 and
additional support from the European Southern Observatory.

\placefigure{figspectra}

\clearpage


\begin{deluxetable}{rrccccccrrrccrl}
\tabcolsep=3pt
\tabletypesize{\scriptsize}
\tablewidth{0pt}
\tablecaption{FIRST Bright Quasar Catalog}
\tablenum{1}
\tablehead{
\colhead{RA}&\colhead{Dec}&\colhead{Cl}&\colhead{R}&\colhead{B}&\colhead{B-R}&\colhead{A(R)}&\colhead{Sep}&\colhead{$S_p$}&\colhead{$S_i$}&\colhead{$M_B$}&\colhead{$\log L_R$}&\colhead{$\log R^*$}&\colhead{z}&\colhead{Notes}\\
\colhead{(1)}&\colhead{(2)}&\colhead{(3)}&\colhead{(4)}&\colhead{(5)}&\colhead{(6)}&\colhead{(7)}&\colhead{(8)}&\colhead{(9)}&\colhead{(10)}&\colhead{(11)}&\colhead{(12)}&\colhead{(13)}&\colhead{(14)}&\colhead{(15)}
}
\startdata
21 02 08.142&-07 33 34.62&* &  18.02&  19.01&  0.99&0.21&   0.72&     1.94&      1.95&-27.6& 32.5&  0.96& 2.660&\\
21 07 57.680&-06 20 10.52&* &  16.64&  18.02&  1.38&0.17&   0.34&    19.21&     19.61&-25.2& 32.3&  1.74& 0.644&\\
21 09 18.684&-07 21 37.99&* &  18.23&  20.15&  1.92&0.30&   0.45&     1.18&      0.69&-24.6& 31.6&  1.30& 1.230&\\
21 09 26.412&-07 39 25.92&* &  18.69&  19.62&  0.93&0.23&   0.80&    44.85&     57.14&-26.1& 33.6&  2.72& 1.880&\\
21 10 13.184&-06 17 20.85&* &  18.28&  19.66&  1.39&0.22&   0.23&    32.89&     36.21&-25.5& 33.2&  2.57& 1.470&\\
21 17 28.905&-06 01 49.02&* &  17.76&  18.20&  0.43&0.35&   0.64&    78.43&     81.04&-27.4& 33.7&  2.31& 1.760&\\
21 18 43.248&-06 36 17.94&* &  16.60&  17.92&  1.32&0.37&   0.39&    75.52&     77.44&-23.7& 32.3&  2.34& 0.330&X\\
21 23 36.041&-01 39 47.38&gR&  18.53&  19.36&  0.83&0.13&   0.19&     1.26&      1.21&-23.2& 30.8&  1.10& 0.494&X\\
\enddata
\end{deluxetable}

\begin{deluxetable}{rrccccccrrrccrl}
\tabcolsep=3pt
\tabletypesize{\scriptsize}
\tablewidth{0pt}
\tablecaption{Narrow-line AGN in FBQS Candidate List}
\tablenum{2}
\tablehead{
\colhead{RA}&\colhead{Dec}&\colhead{Cl}&\colhead{R}&\colhead{B}&\colhead{B-R}&\colhead{A(R)}&\colhead{Sep}&\colhead{$S_p$}&\colhead{$S_i$}&\colhead{$M_B$}&\colhead{$\log L_R$}&\colhead{$\log R^*$}&\colhead{z}&\colhead{Notes}\\
\colhead{(1)}&\colhead{(2)}&\colhead{(3)}&\colhead{(4)}&\colhead{(5)}&\colhead{(6)}&\colhead{(7)}&\colhead{(8)}&\colhead{(9)}&\colhead{(10)}&\colhead{(11)}&\colhead{(12)}&\colhead{(13)}&\colhead{(14)}&\colhead{(15)}
}
\startdata
21 17 05.992&-06 40 46.07&gR&  17.91&  19.51&  1.60&0.54&   0.24&     2.06&      1.71&-20.8& 30.2&  1.42& 0.185&\\
21 31 52.555&-01 01 52.02&* &  18.44&  20.22&  1.78&0.14&   0.14&     4.09&      4.41&-21.8& 31.2&  2.00& 0.390&I\\
21 39 50.995&-08 04 54.67&gR&  12.64&  13.39&  0.75&0.11&   0.39&     5.83&     11.41&-24.1& 29.8& -0.25& 0.051&I\\
21 45 08.153&-00 23 41.21&gR&  18.68&  20.64&  1.96&0.18&   0.39&     1.70&      1.85&-20.1& 30.3&  1.82& 0.226&\\
22 06 02.534&-08 21 06.44&gR&  11.82&  12.34&  0.52&0.09&   1.20&     2.33&      2.88&-25.7& 29.5& -1.28& 0.067&\\
22 23 40.975&-08 24 57.95&gB&  18.76&  20.41&  1.65&0.14&   0.08&     5.53&      5.37&-21.2& 31.1&  2.19& 0.330&\\
22 27 07.151&-00 18 36.05&* &  17.64&  18.57&  0.93&0.24&   0.44&     1.03&      1.81&-21.0& 29.8&  1.00& 0.130&\\
22 51 22.146&-08 57 23.06&gR&  16.69&  17.72&  1.03&0.12&   0.82&     1.47&      1.67&-20.7& 29.4&  0.64& 0.080&\\
\enddata
%
\tablecomments{Table columns are same as Table 1.}
\end{deluxetable}

\begin{deluxetable}{rrccccccrrrccrl}
\tabcolsep=3pt
\tabletypesize{\scriptsize}
\tablewidth{0pt}
\tablecaption{BL Lacertae Objects in FBQS Candidate List}
\tablenum{3}
\tablehead{
\colhead{RA}&\colhead{Dec}&\colhead{Cl}&\colhead{R}&\colhead{B}&\colhead{B-R}&\colhead{A(R)}&\colhead{Sep}&\colhead{$S_p$}&\colhead{$S_i$}&\colhead{$M_B$}&\colhead{$\log L_R$}&\colhead{$\log R^*$}&\colhead{z}&\colhead{Notes}\\
\colhead{(1)}&\colhead{(2)}&\colhead{(3)}&\colhead{(4)}&\colhead{(5)}&\colhead{(6)}&\colhead{(7)}&\colhead{(8)}&\colhead{(9)}&\colhead{(10)}&\colhead{(11)}&\colhead{(12)}&\colhead{(13)}&\colhead{(14)}&\colhead{(15)}
}
\startdata
21 29 40.691&+00 35 27.86&gR&  18.81&  20.00&  1.18&0.12&   0.25&     5.86&      5.92&-22.2& 31.4&  2.04& 0.425&\\
21 34 10.335&-01 53 17.39&* &  18.25&  19.83&  1.58&0.15&   0.32&  1498.44&   1619.87&\nodata&\nodata&  4.49&\nodata&o\\
21 56 14.755&-00 37 04.58&* &  18.62&  20.51&  1.89&0.32&   0.19&   165.65&    170.17&\nodata&\nodata&  3.78&\nodata&\\
21 56 50.339&-08 55 35.36&* &  18.24&  19.73&  1.49&0.09&   0.19&    20.45&     21.49&\nodata&\nodata&  2.57&\nodata&\\
22 04 45.256&+00 31 42.32&* &  16.73&  18.35&  1.62&0.15&   0.61&     2.94&      3.03&-26.6& 32.1&  0.98& 1.330&\\
22 11 08.341&-00 03 02.52&* &  17.53&  18.91&  1.38&0.17&   0.16&    15.25&     16.89&\nodata&\nodata&  2.14&\nodata&X\\
22 17 25.998&-09 07 05.21&gR&  12.76&  13.15&  0.39&0.19&   0.79&     1.11&      3.73&-25.7& 29.9& -0.85& 0.095&\\
22 25 58.839&-01 45 12.07&gR&  18.08&  18.73&  0.65&0.13&   0.17&     8.75&      9.01&\nodata&\nodata&  1.79&\nodata&?\\
\enddata
%
\tablecomments{Table columns are the same as in Table 1.}
\end{deluxetable}

\begin{deluxetable}{rrccccccrrrccrl}
\tabcolsep=3pt
\tabletypesize{\scriptsize}
\tablewidth{0pt}
\tablecaption{H II Region Galaxies in FBQS Candidate List}
\tablenum{4}
\tablehead{
\colhead{RA}&\colhead{Dec}&\colhead{Cl}&\colhead{R}&\colhead{B}&\colhead{B-R}&\colhead{A(R)}&\colhead{Sep}&\colhead{$S_p$}&\colhead{$S_i$}&\colhead{$M_B$}&\colhead{$\log L_R$}&\colhead{$\log R^*$}&\colhead{z}&\colhead{Notes}\\
\colhead{(1)}&\colhead{(2)}&\colhead{(3)}&\colhead{(4)}&\colhead{(5)}&\colhead{(6)}&\colhead{(7)}&\colhead{(8)}&\colhead{(9)}&\colhead{(10)}&\colhead{(11)}&\colhead{(12)}&\colhead{(13)}&\colhead{(14)}&\colhead{(15)}
}
\startdata
21 22 20.646&-00 00 46.38&gR&  18.28&  19.86&  1.59&0.15&   0.75&     1.48&      0.75&-21.5& 30.5&  1.40& 0.297&\\
21 41 55.403&-08 49 15.63&gR&  17.46&  18.71&  1.25&0.10&   0.13&     1.04&      0.78&-22.0& 30.1&  0.80& 0.220&\\
22 48 14.954&-01 09 23.14&gR&  18.38&  20.23&  1.84&0.27&   0.69&     1.14&      1.11&-19.3& 29.6&  1.46& 0.128&\\
23 27 14.545&-10 23 17.75&gR&  12.79&  12.57& -0.22&0.08&   0.30&     1.41&      2.25&-25.4& 29.3& -1.29& 0.065&I\\
23 30 54.903&-00 00 40.14&gR&  18.53&  20.38&  1.85&0.12&   0.12&     2.22&      2.68&-19.8& 30.3&  1.89& 0.174&\\
23 56 59.633&-02 05 00.85&* &  12.04&  12.84&  0.80&0.10&   0.15&     1.05&      1.23&-23.0& 28.2& -1.43& 0.024&I\\
00 06 18.889&-08 53 33.05&gR&  13.44&  13.09& -0.35&0.10&   0.95&     1.06&      1.98&-25.0& 29.3& -1.14& 0.068&\\
00 06 39.951&-01 31 40.48&gB&  18.12&  19.28&  1.16&0.11&   0.29&     3.43&      3.43&-21.1& 30.5&  1.55& 0.193&\\
\enddata
%
\tablecomments{Table columns are same as Table 1.}
\end{deluxetable}

\begin{deluxetable}{rrccccccrrrccrl}
\tabcolsep=3pt
\tabletypesize{\scriptsize}
\tablewidth{0pt}
\tablecaption{Passive Galaxies in FBQS Candidate List}
\tablenum{5}
\tablehead{
\colhead{RA}&\colhead{Dec}&\colhead{Cl}&\colhead{R}&\colhead{B}&\colhead{B-R}&\colhead{A(R)}&\colhead{Sep}&\colhead{$S_p$}&\colhead{$S_i$}&\colhead{$M_B$}&\colhead{$\log L_R$}&\colhead{$\log R^*$}&\colhead{z}&\colhead{Notes}\\
\colhead{(1)}&\colhead{(2)}&\colhead{(3)}&\colhead{(4)}&\colhead{(5)}&\colhead{(6)}&\colhead{(7)}&\colhead{(8)}&\colhead{(9)}&\colhead{(10)}&\colhead{(11)}&\colhead{(12)}&\colhead{(13)}&\colhead{(14)}&\colhead{(15)}
}
\startdata
21 20 30.602&-01 38 40.45&gR&  18.75&  19.99&  1.23&0.13&   0.23&     6.30&      6.27&-21.7& 31.2&  2.07& 0.334&\\
22 26 02.367&+00 09 30.81&gR&  18.38&  20.23&  1.85&0.22&   0.05&     2.13&      1.59&-21.1& 30.6&  1.71& 0.290&\\
22 26 34.074&-09 01 06.47&* &  18.13&  19.43&  1.30&0.13&   0.60&     1.73&      1.72&-22.0& 30.5&  1.30& 0.299&I\\
23 18 49.658&-09 35 15.52&gR&  18.85&  20.56&  1.71&0.08&   0.46&     4.07&      3.93&-20.3& 30.7&  2.13& 0.236&\\
23 42 03.314&-08 51 28.17&gR&  18.84&  20.55&  1.71&0.08&   0.17&     1.88&      2.85&\nodata&\nodata&  2.02&\nodata&o\\
23 45 57.122&-10 28 48.20&gR&  17.86&  18.81&  0.95&0.09&   0.42&    11.49&     12.39&-22.8& 31.5&  1.90& 0.330&\\
23 47 12.620&-10 34 00.38&gR&  18.66&  20.17&  1.51&0.09&   0.33&     2.88&      2.91&-21.2& 30.7&  1.82& 0.289&\\
23 50 48.109&-11 21 43.78&gR&  17.38&  19.10&  1.72&0.07&   0.82&     1.82&      1.82&-22.3& 30.6&  1.19& 0.302&\\
\enddata
%
\tablecomments{Table columns are same as Table 1.}
\end{deluxetable}

\begin{deluxetable}{rrccccccrrl}
\tabcolsep=3pt
\tabletypesize{\scriptsize}
\tablewidth{0pt}
\tablecaption{Stars in FBQS Candidate List}
\tablenum{6}
\tablehead{
\colhead{RA}&\colhead{Dec}&\colhead{Cl}&\colhead{R}&\colhead{B}&\colhead{B-R}&\colhead{A(R)}&\colhead{Sep}&\colhead{$S_p$}&\colhead{$S_i$}&\colhead{Notes}\\
\colhead{(1)}&\colhead{(2)}&\colhead{(3)}&\colhead{(4)}&\colhead{(5)}&\colhead{(6)}&\colhead{(7)}&\colhead{(8)}&\colhead{(9)}&\colhead{(10)}&\colhead{(11)}
}
\startdata
21 03 36.181&-07 15 34.78&* &  17.63&  18.94&  1.31&0.25&   0.93&     1.32&      0.83&\\
21 12 27.627&-07 19 31.73&* &  17.98&  19.58&  1.60&0.40&   0.33&     1.12&     10.29&\\
21 13 57.288&-06 41 35.13&* &  11.74&  12.51&  0.77&0.33&   0.70&     1.15&      1.44&\\
21 16 11.898&-06 28 30.09&gR&  17.74&  18.85&  1.11&0.37&   0.44&    13.79&     14.87&\\
21 17 58.907&-07 46 11.55&* &  14.19&  15.25&  1.05&0.39&   0.90&     3.16&      3.39&\\
21 29 26.165&+00 24 35.38&* &  15.36&  17.07&  1.70&0.12&   0.09&     4.29&      4.46&\\
21 31 57.191&+01 30 25.51&* &  16.72&  18.51&  1.79&0.14&   0.67&     6.51&     11.17&\\
21 35 20.147&-01 04 26.03&* &  18.60&  19.73&  1.13&0.13&   1.37&     2.66&      3.53&CJ\\
\enddata
%
\tablecomments{Table columns are same as Table 1, except
that redshift, absolute magnitudes, etc., are omitted for
these Galactic objects.}
\end{deluxetable}

\begin{deluxetable}{rrccccccrrcl}
\tabcolsep=3pt
\tabletypesize{\scriptsize}
\tablewidth{0pt}
\tablecaption{Objects Without Spectra in FBQS Candidate List}
\tablenum{7}
\tablehead{
\colhead{RA}&\colhead{Dec}&\colhead{Cl}&\colhead{R}&\colhead{B}&\colhead{B-R}&\colhead{A(R)}&\colhead{Sep}&\colhead{$S_p$}&\colhead{$S_i$}&\colhead{P(Q)}&\colhead{Notes}\\
\colhead{(1)}&\colhead{(2)}&\colhead{(3)}&\colhead{(4)}&\colhead{(5)}&\colhead{(6)}&\colhead{(7)}&\colhead{(8)}&\colhead{(9)}&\colhead{(10)}&\colhead{(11)}&\colhead{(12)}
}
\startdata
21 26 27.266&+01 23 21.44&* &  18.86&  19.51&  0.65&0.12&   0.83&     1.37&      1.74&0.87&\\
21 57 55.635&-07 46 53.17&* &  17.87&  18.78&  0.91&0.11&   0.56&     1.08&     17.33&0.14&\\
22 09 08.252&-00 55 59.06&gR&  18.61&  19.90&  1.29&0.27&   0.32&    20.88&     21.16&0.65&\\
22 11 09.861&-00 23 27.20&gR&  18.42&  20.26&  1.84&0.30&   0.65&    29.31&     38.84&0.51&\\
22 20 35.227&-07 49 47.34&gB&  17.79&  17.11& -0.68&0.20&   0.19&    10.46&     12.92&0.42&\\
22 24 46.000&-08 46 56.42&* &  18.84&  19.38&  0.55&0.13&   0.11&    34.93&     36.36&0.88&\\
22 26 52.940&-00 07 15.07&gR&  18.02&  19.63&  1.61&0.21&   1.09&     1.90&      1.51&0.27&\\
22 35 16.241&-00 58 49.77&* &  18.85&  19.43&  0.58&0.14&   0.04&     2.91&      2.28&0.88&\\
\enddata
%
\tablecomments{Table columns are same as Table 6, with the addition
of Col.~11. $P(Q)$ is the probability that the candidate is
a quasar, computed as described in Paper II.
All source parameters including color were used.}
\end{deluxetable}

\begin{deluxetable}{rrrccrcl}
\tabcolsep=3pt
\tablewidth{0pt}
\tablecaption{Broad Absorption-Line Quasars}
\tablenum{8}
\tablehead{
\colhead{RA}&\colhead{Dec}&\colhead{$M_B$}&\colhead{$\log L_R$}&\colhead{$\log R^*$}&\colhead{z}&\colhead{BALnicity\tablenotemark{a}}&\colhead{Notes}\\
\colhead{(1)}&\colhead{(2)}&\colhead{(3)}&\colhead{(4)}&\colhead{(5)}&\colhead{(6)}&\colhead{(7)}&\colhead{(8)}
}
\startdata
21 24 10.271&-07 22 19.95&-27.7& 33.0&  1.49& 1.760&\phn800&HiBAL\\
23 00 11.749&-10 21 44.02&-27.7& 34.3&  2.74& 2.300&\phn\phn\phn0&HiBAL?\\
23 36 11.627&-09 54 27.05&-27.1& 32.1&  0.80& 1.760&\phn100&HiBAL\\
23 47 11.497&-10 37 42.18&-28.0& 32.0&  0.33& 1.780&9000&LoBAL\\
00 14 38.279&-01 07 50.31&-25.7& 32.0&  1.30& 1.800&\phn\phn\phn0&LoBAL?\\
00 46 13.548&+01 04 25.78&-28.0& 32.5&  0.80& 2.150&2100&o,HiBAL\\
01 28 41.876&-00 33 16.79&-26.2& 32.2&  1.31& 1.650&2700&HiBAL\\
01 35 15.239&-02 13 49.26&-28.0& 33.2&  1.56& 1.820&1900&HiBAL\\
01 40 18.199&-01 38 05.94&-28.2& 32.2&  0.46& 2.240&1400&o,HiBAL\\
01 48 12.829&-00 51 09.01&-26.5& 32.4&  1.32& 1.820&6800&HiBAL\\
01 53 45.346&-10 08 40.06&-26.5& 32.5&  1.46& 1.900&\phn600&HiBAL\\
02 00 22.031&-08 45 11.76&-26.2& 32.8&  1.85& 1.930&2300&HiBAL\\
02 21 00.743&-01 22 12.16&-25.5& 31.8&  1.10& 1.210&\phn\phn50\tablenotemark{b}&LoBAL\\
02 56 25.641&-01 19 11.51&-27.7& 33.5&  1.99& 2.490&\phn250&HiBAL
\tablenotetext{a}{Table columns 1--6 and 8 are from Table 1. Column 7 is
the ``BALnicity index'' \citep{weymann91}, which if non-zero indicates
high confidence that the absorption is a true broad-absorption line
rather than associated absorption.}
\tablenotetext{b}{BALnicity was computed from Mg~II, which generally is narrower
than the high ionization lines.}
\enddata
\end{deluxetable}


\clearpage

\begin{figure*}
\plotone{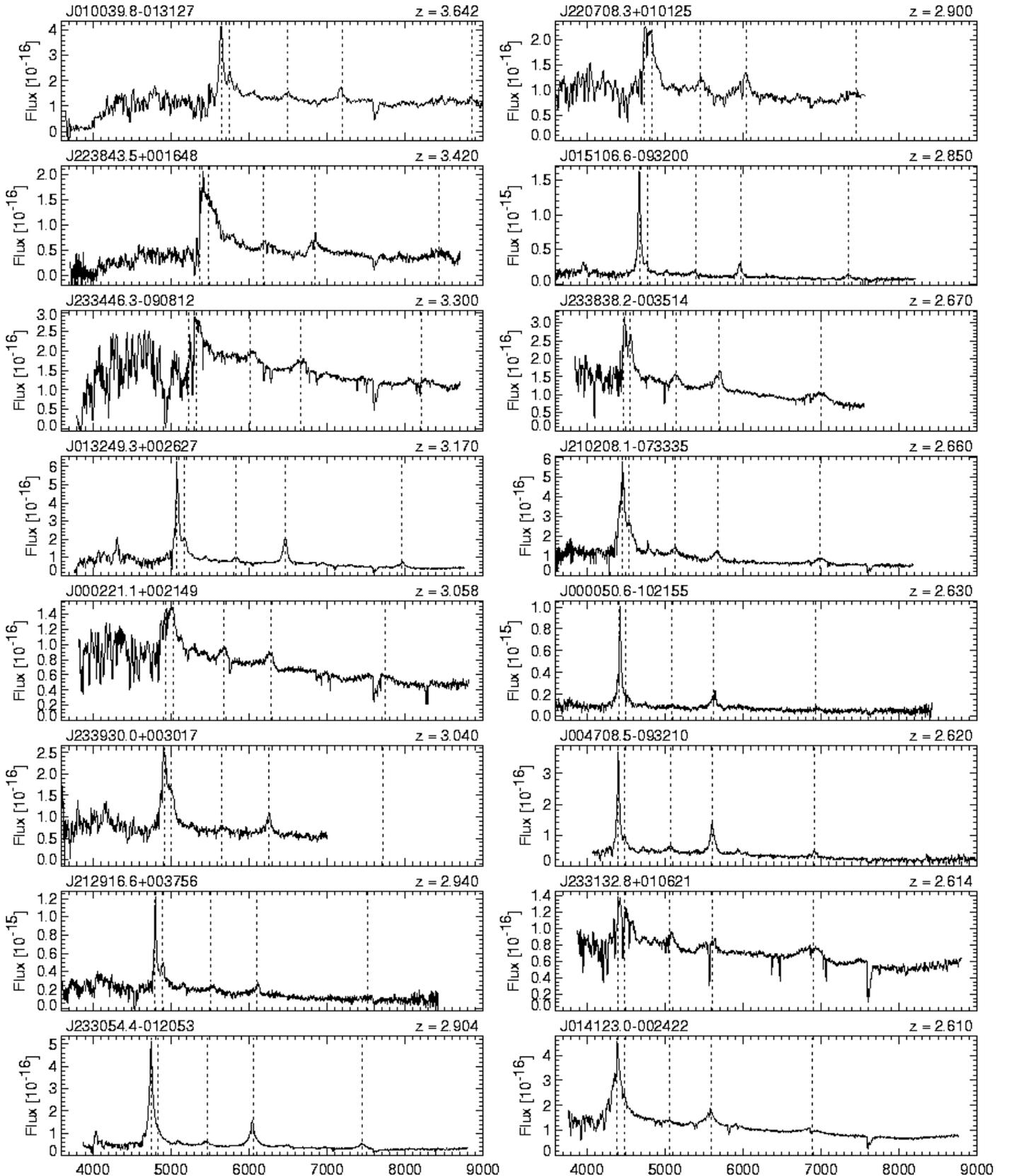}
\caption{
Spectra of FBQS candidates identified as quasars, sorted by
decreasing redshift.
The dotted lines show expected
positions of prominent emission lines:
Ly$\alpha$~1216,
N~V~1240,
Si~IV~1400,
C~IV~1550,
C~III]~1909,
Mg~II~2800,
H$\delta$~4102,
H$\gamma$~4341,
H$\beta$~4862,
[O~III]~4959,
[O~III]~5007,
H$\alpha$~6563.
Note that most of the spectra have atmospheric
A and B band absorption at $\sim6880$~\AA\ and 7620~\AA.
\label{figspectra}
}
\end{figure*}
\clearpage

\begin{figure*}[p]
\figurenum{9b}
\plotone{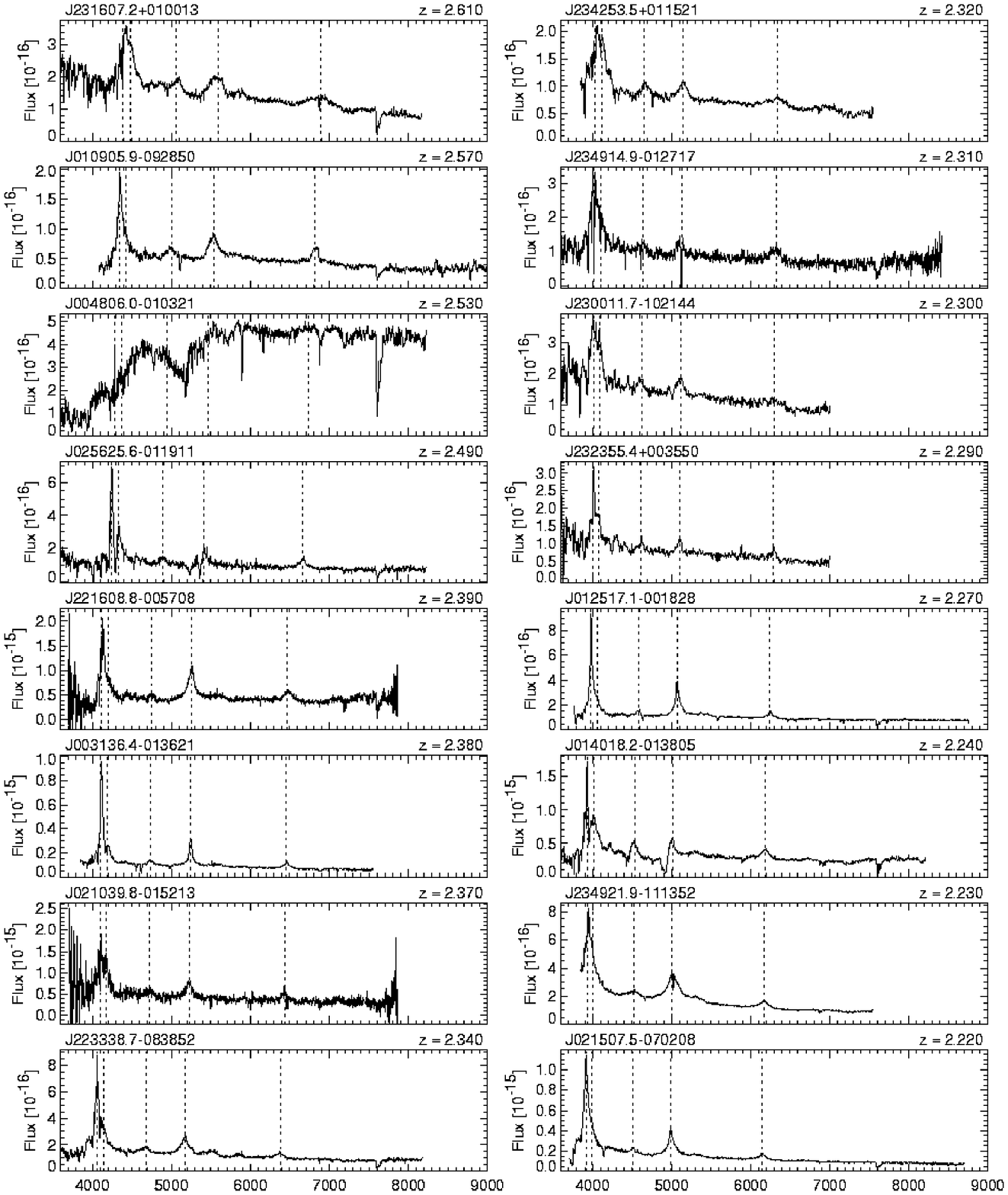}
\caption{{\it Continued.}  Spectra of FBQS quasars.}
\end{figure*}
\clearpage

\begin{figure*}[p]
\figurenum{9c}
\plotone{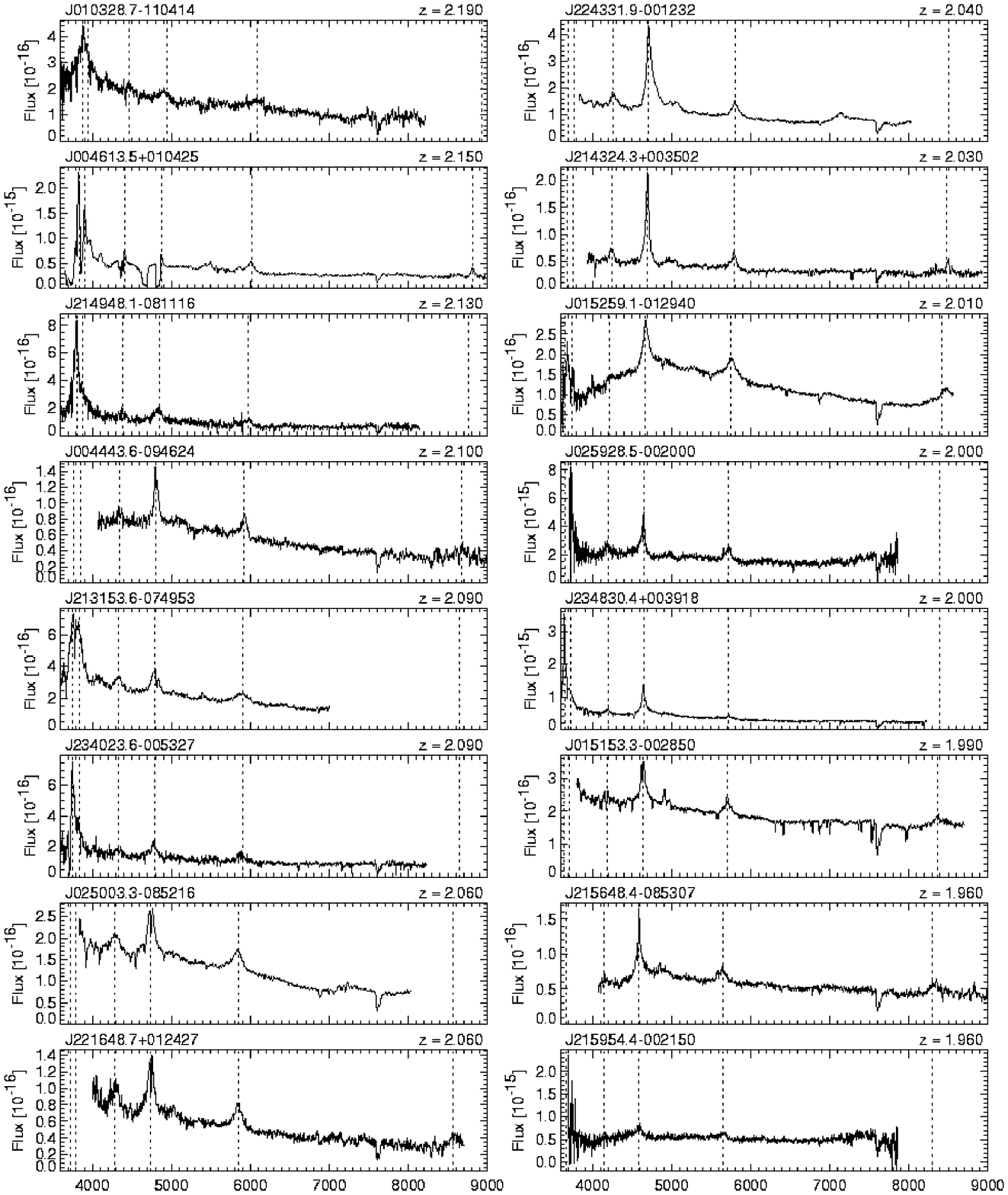}
\caption{{\it Continued.}  Spectra of FBQS quasars.}
\end{figure*}
\clearpage

\begin{figure*}[p]
\figurenum{9d}
\plotone{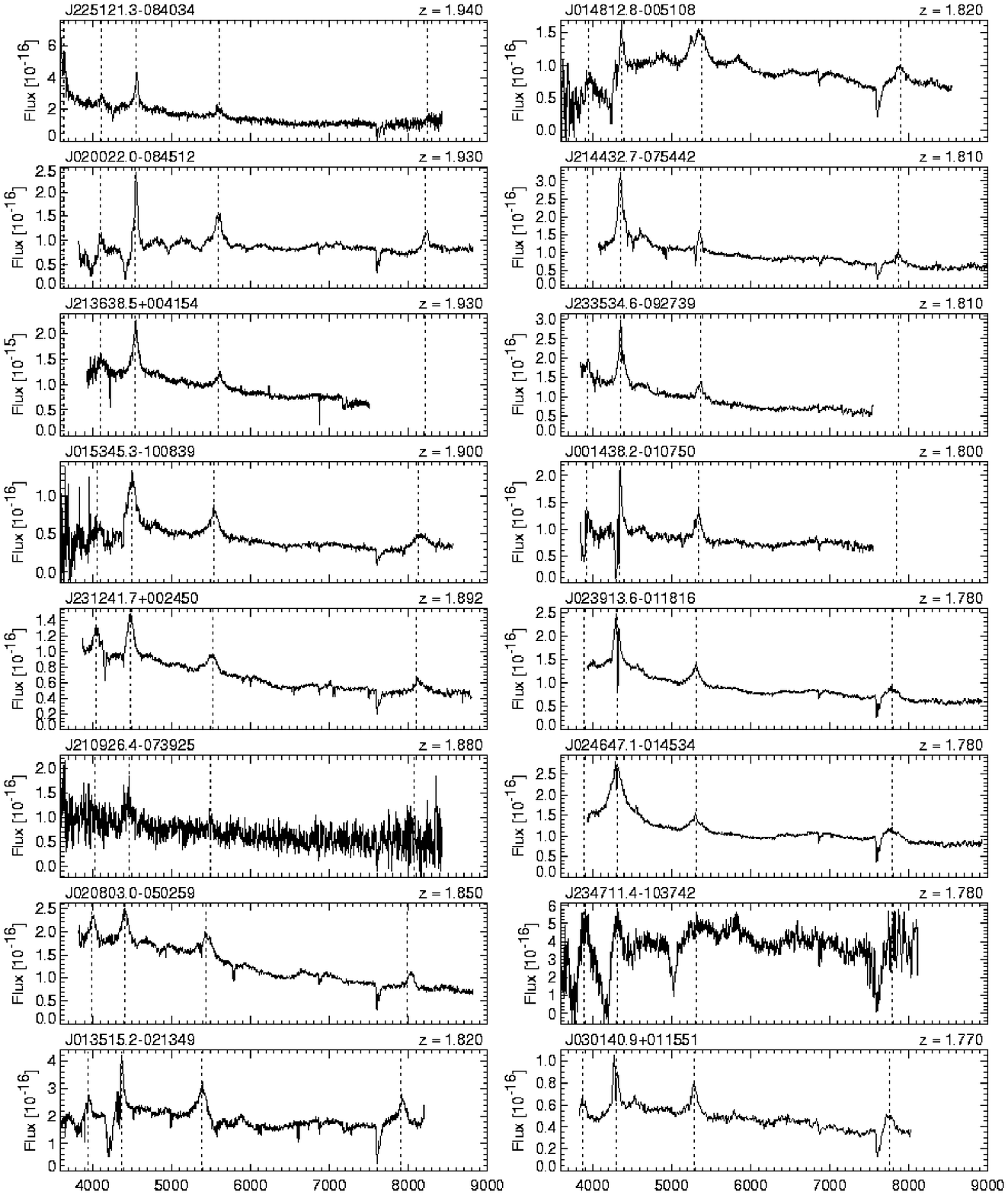}
\caption{{\it Continued.}  Spectra of FBQS quasars.}
\end{figure*}
\clearpage

\begin{figure*}[p]
\figurenum{9e}
\plotone{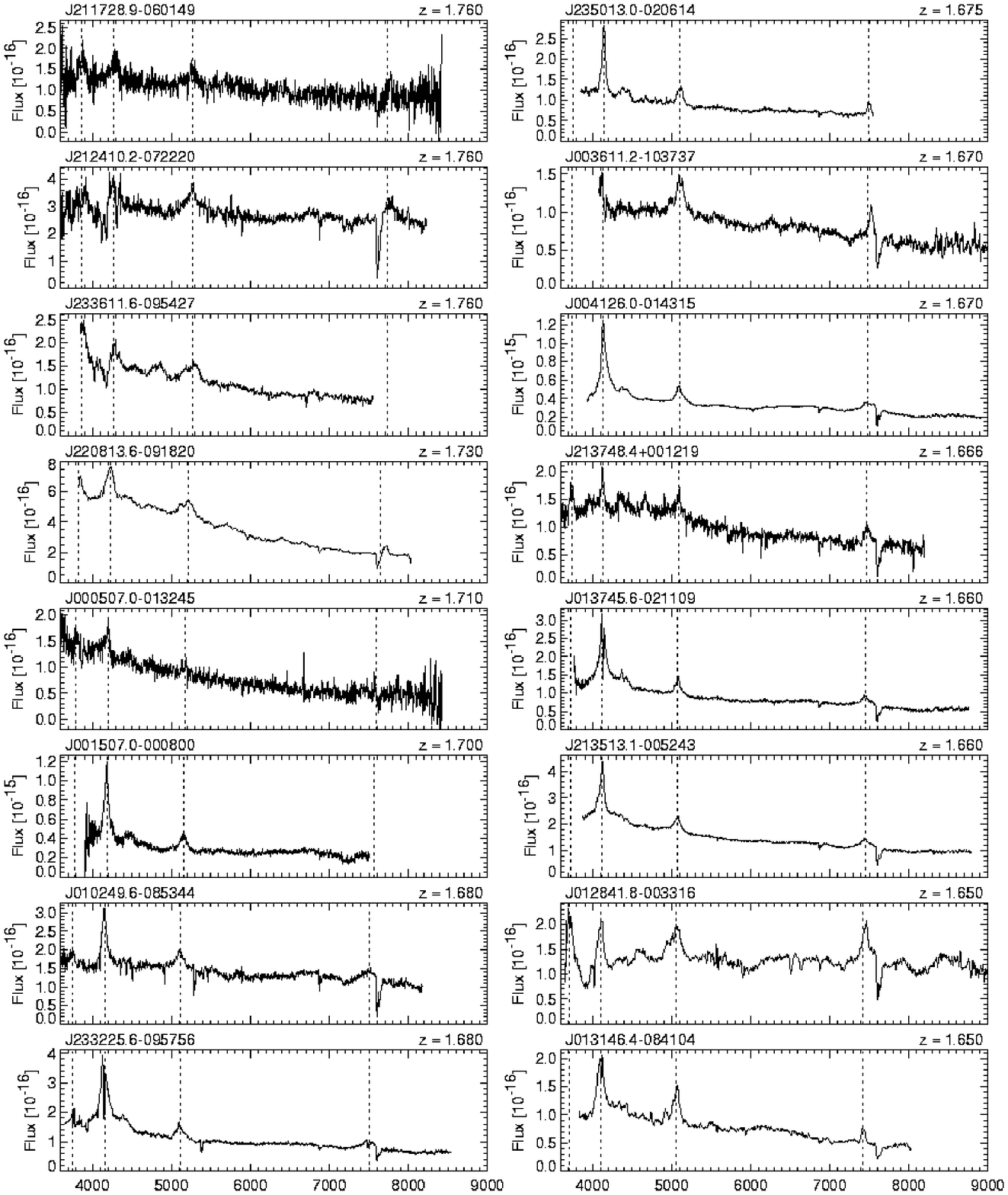}
\caption{{\it Continued.}  Spectra of FBQS quasars.}
\end{figure*}
\clearpage

\begin{figure*}[p]
\figurenum{9f}
\plotone{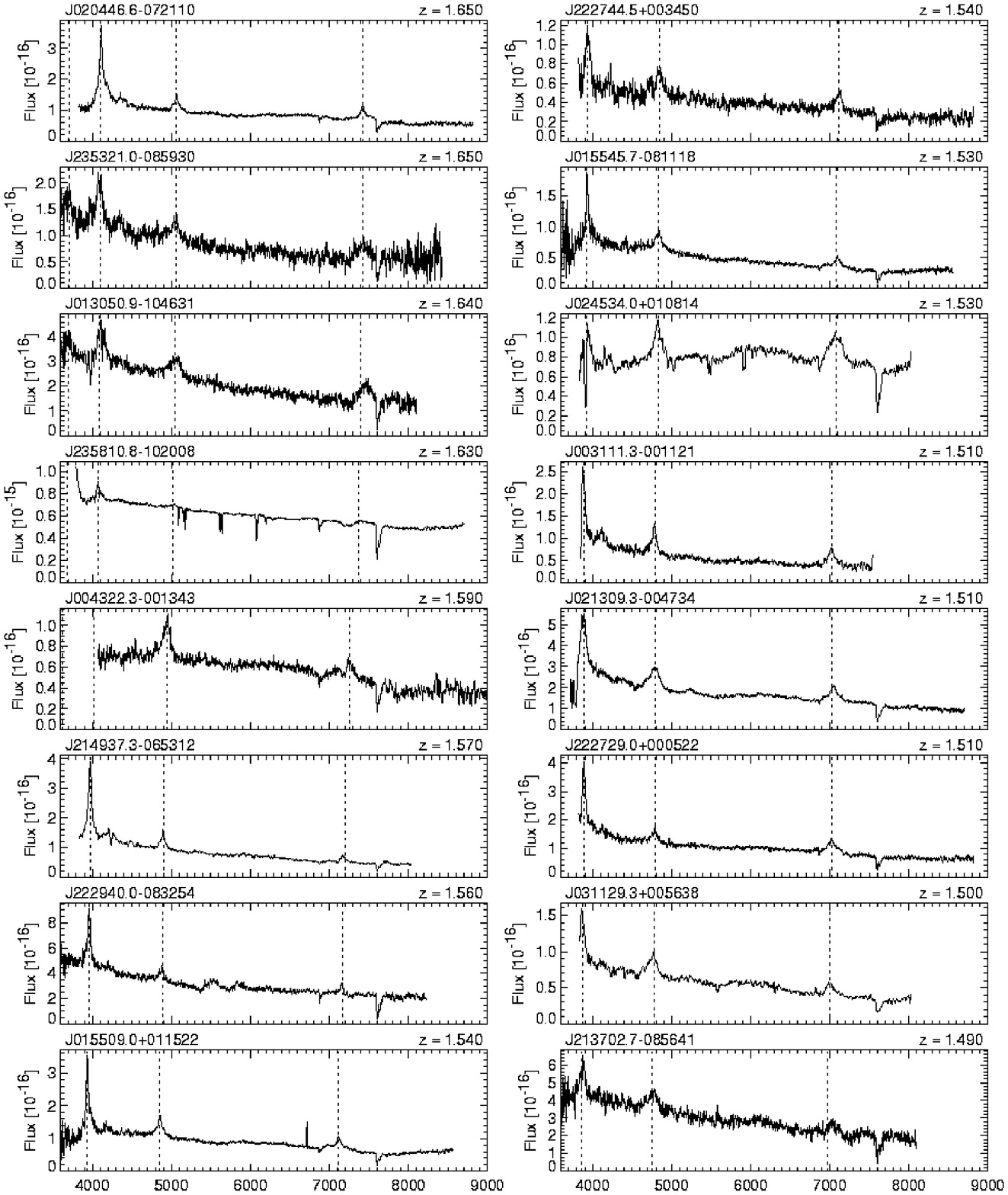}
\caption{{\it Continued.}  Spectra of FBQS quasars.}
\end{figure*}
\clearpage

\begin{figure*}[p]
\figurenum{9g}
\plotone{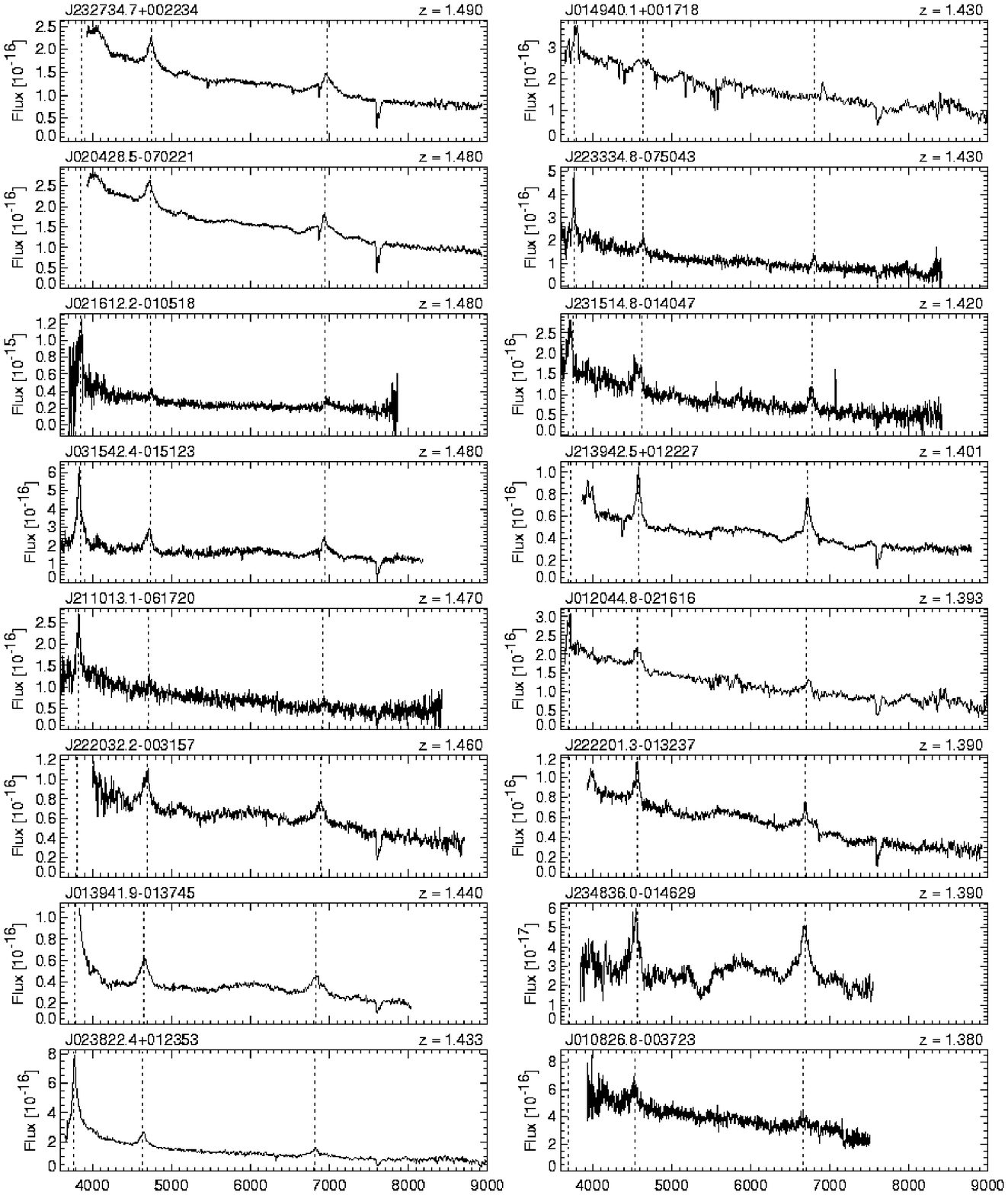}
\caption{{\it Continued.}  Spectra of FBQS quasars.}
\end{figure*}
\clearpage

\begin{figure*}[p]
\figurenum{9h}
\plotone{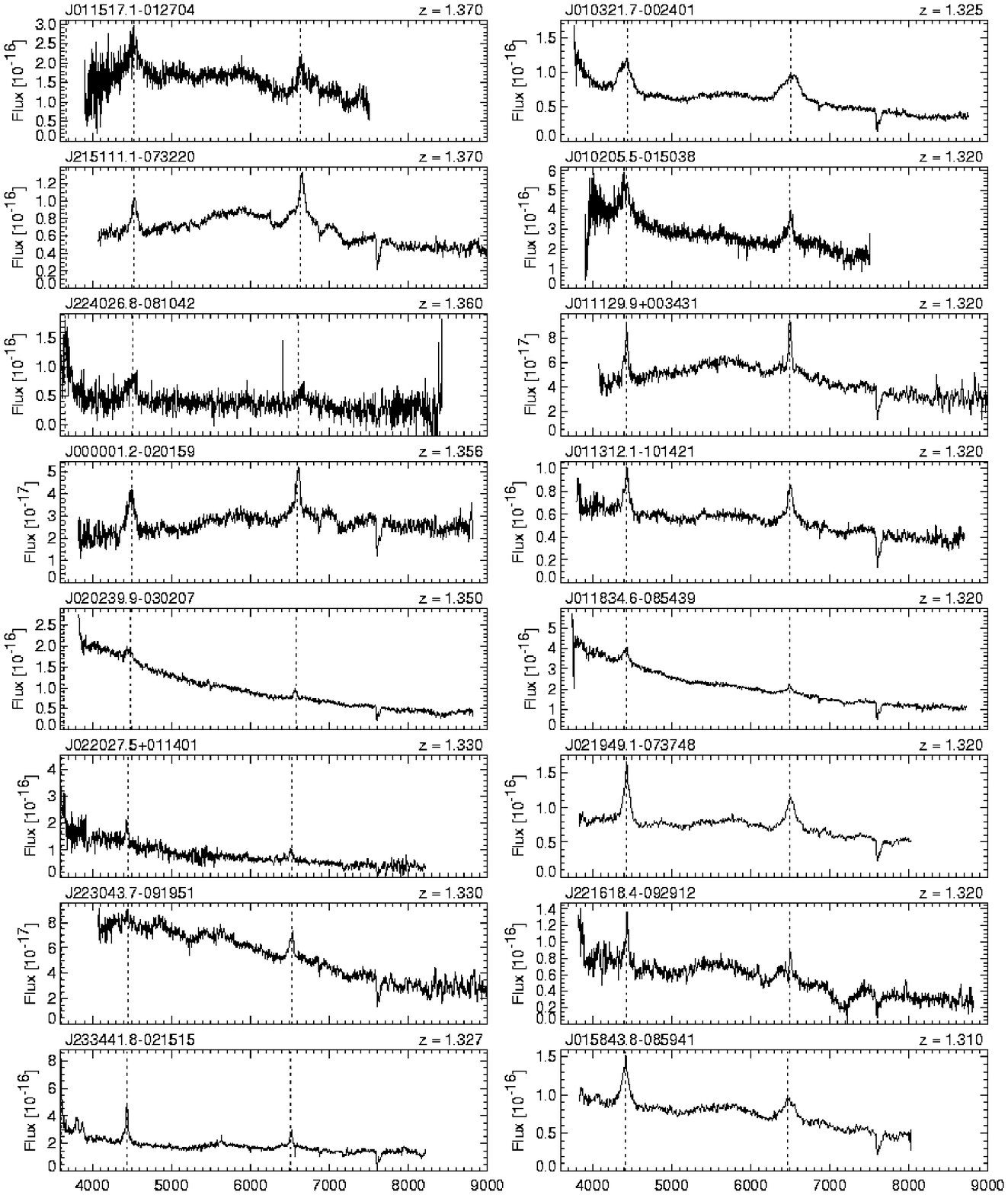}
\caption{{\it Continued.}  Spectra of FBQS quasars.}
\end{figure*}
\clearpage

\begin{figure*}[p]
\figurenum{9i}
\plotone{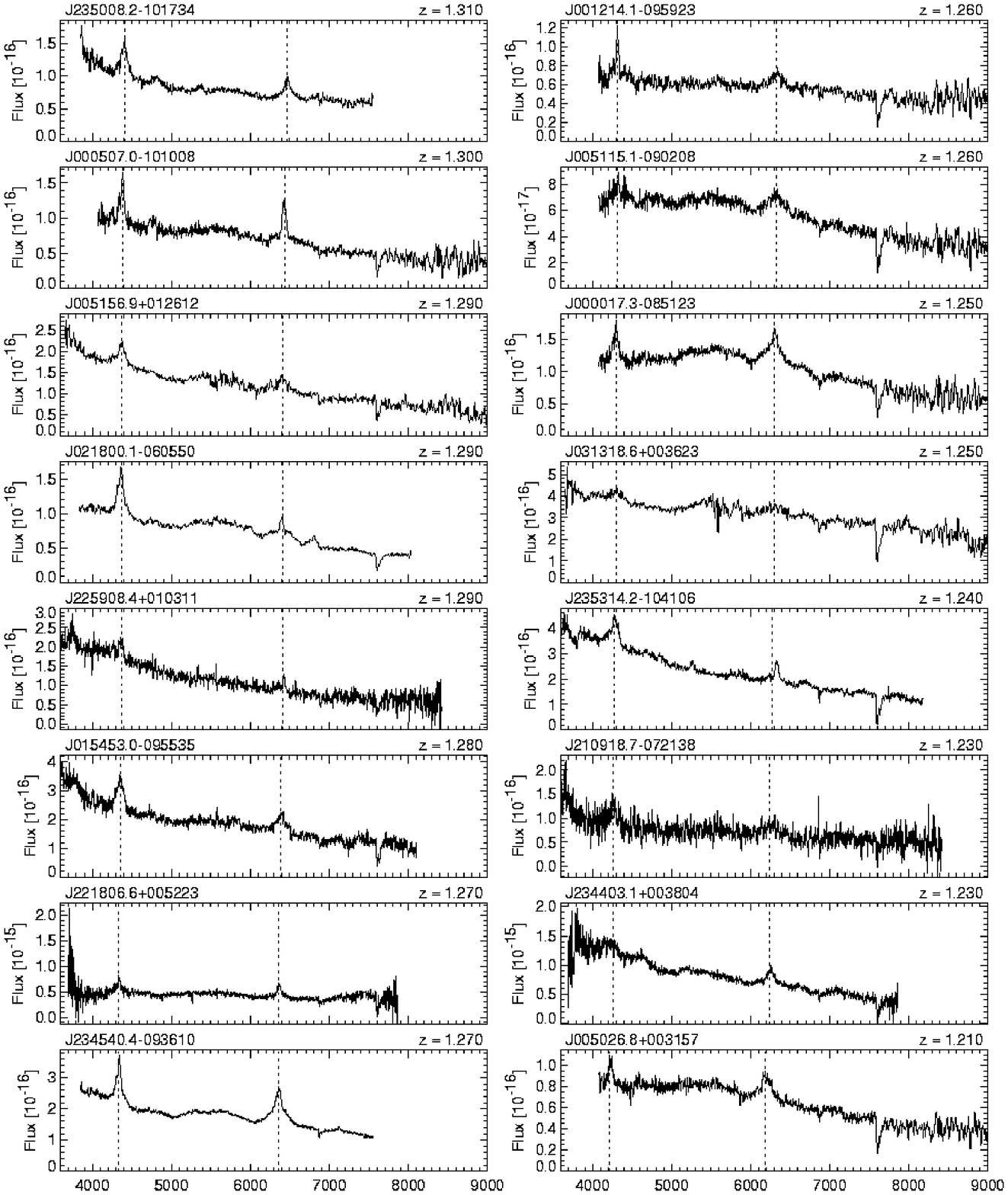}
\caption{{\it Continued.}  Spectra of FBQS quasars.}
\end{figure*}
\clearpage

\begin{figure*}[p]
\figurenum{9j}
\plotone{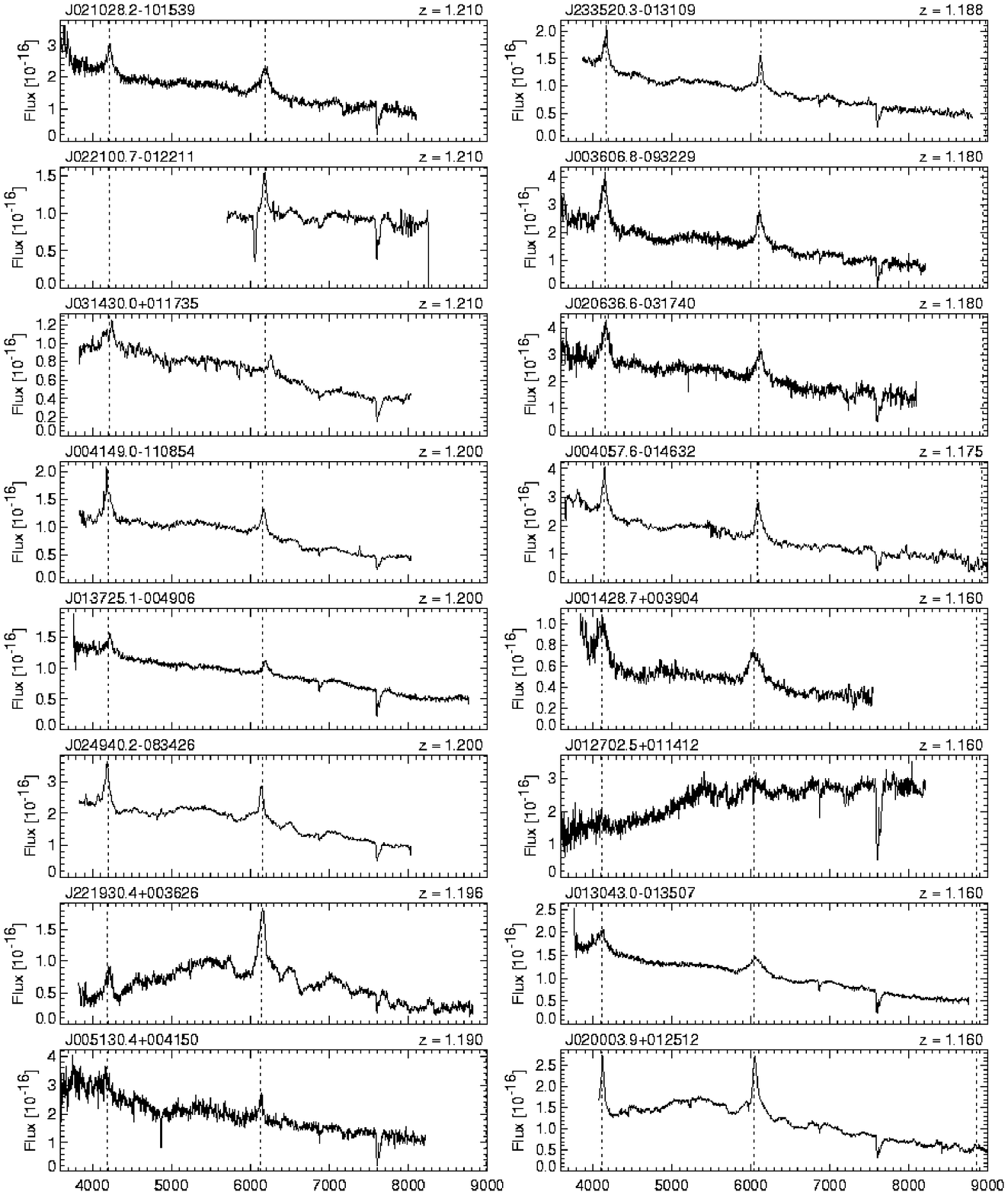}
\caption{{\it Continued.}  Spectra of FBQS quasars.}
\end{figure*}
\clearpage

\begin{figure*}[p]
\figurenum{9k}
\plotone{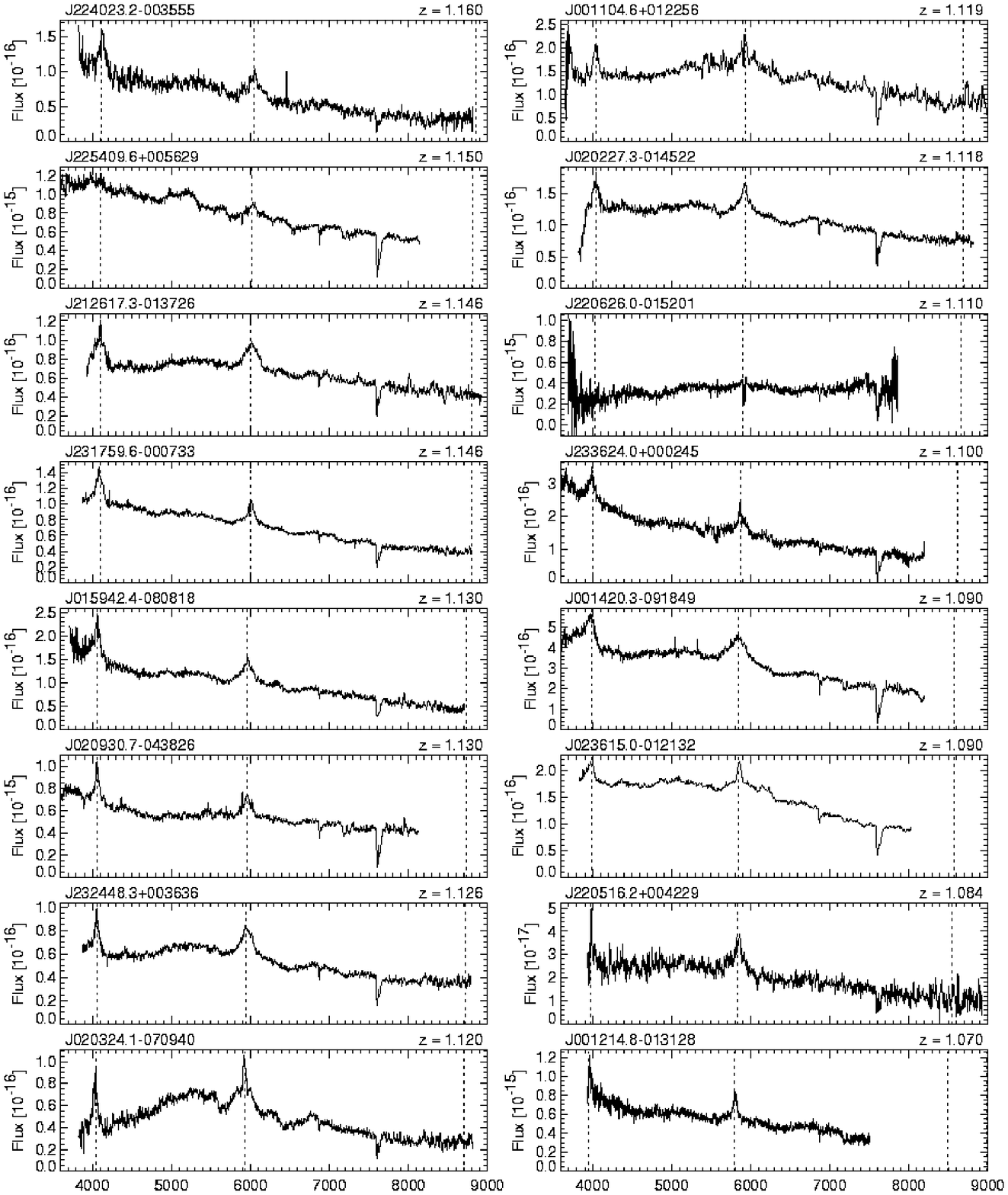}
\caption{{\it Continued.}  Spectra of FBQS quasars.}
\end{figure*}
\clearpage

\begin{figure*}[p]
\figurenum{9l}
\plotone{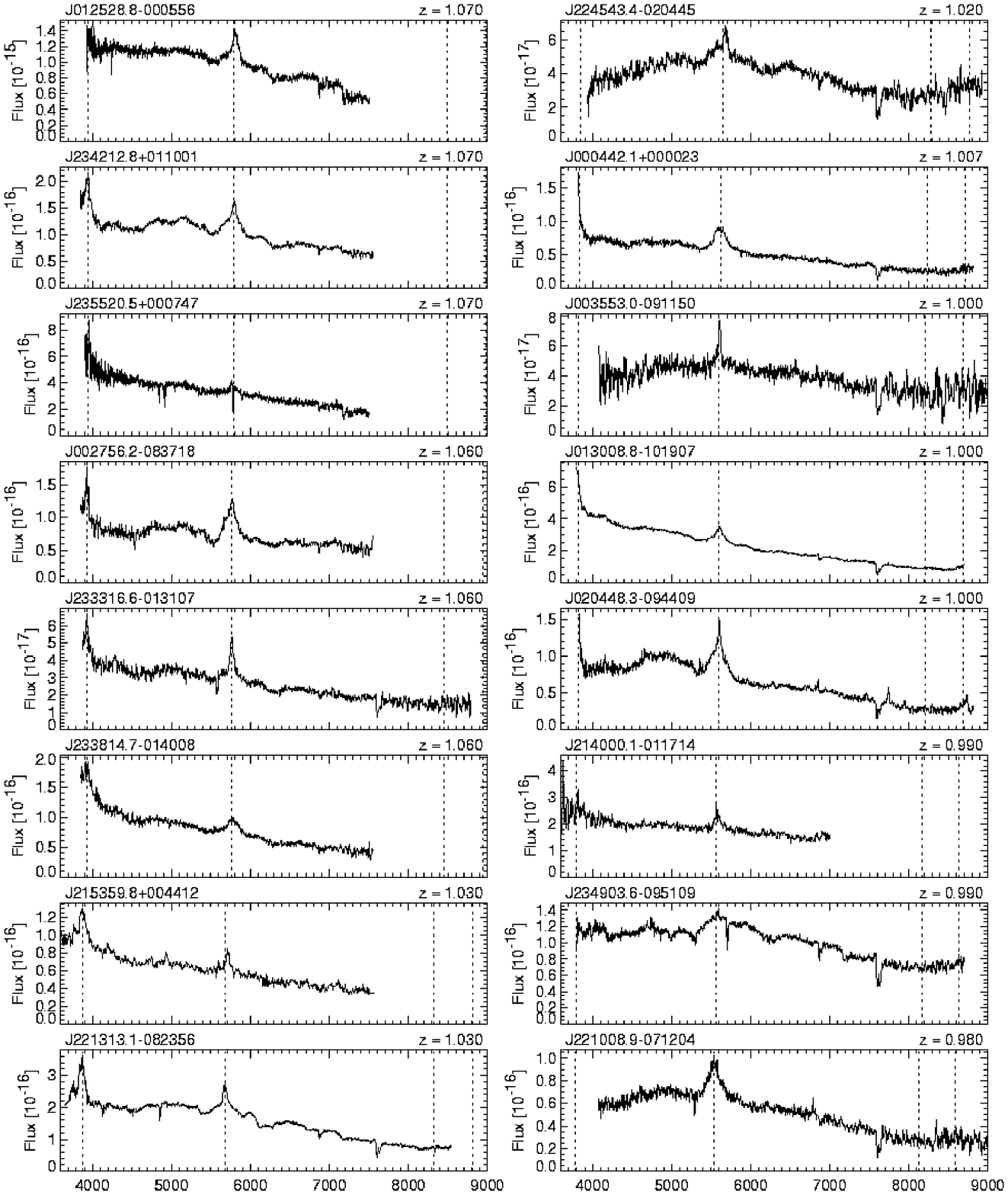}
\caption{{\it Continued.}  Spectra of FBQS quasars.}
\end{figure*}
\clearpage

\begin{figure*}[p]
\figurenum{9m}
\plotone{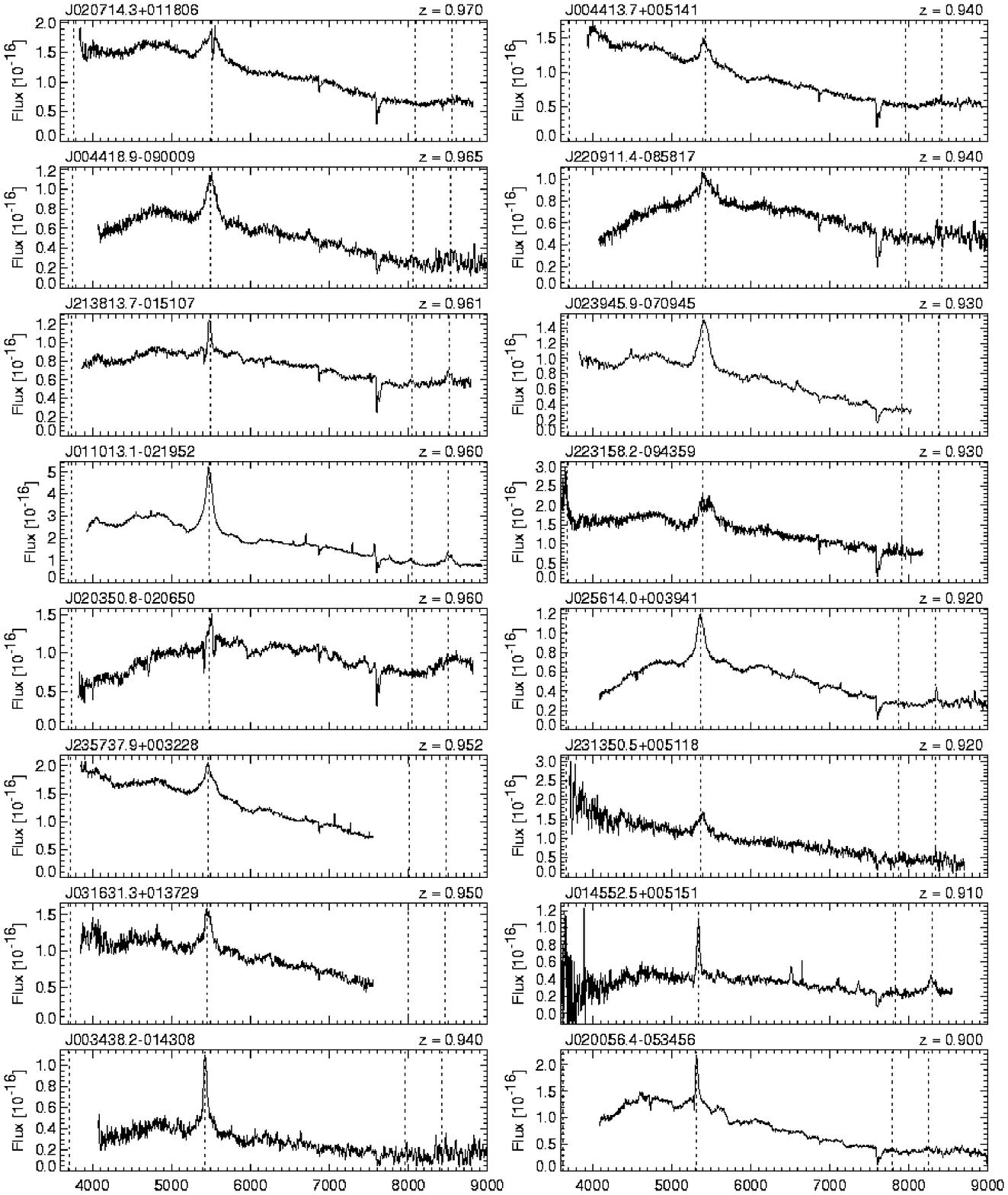}
\caption{{\it Continued.}  Spectra of FBQS quasars.}
\end{figure*}
\clearpage

\begin{figure*}[p]
\figurenum{9n}
\plotone{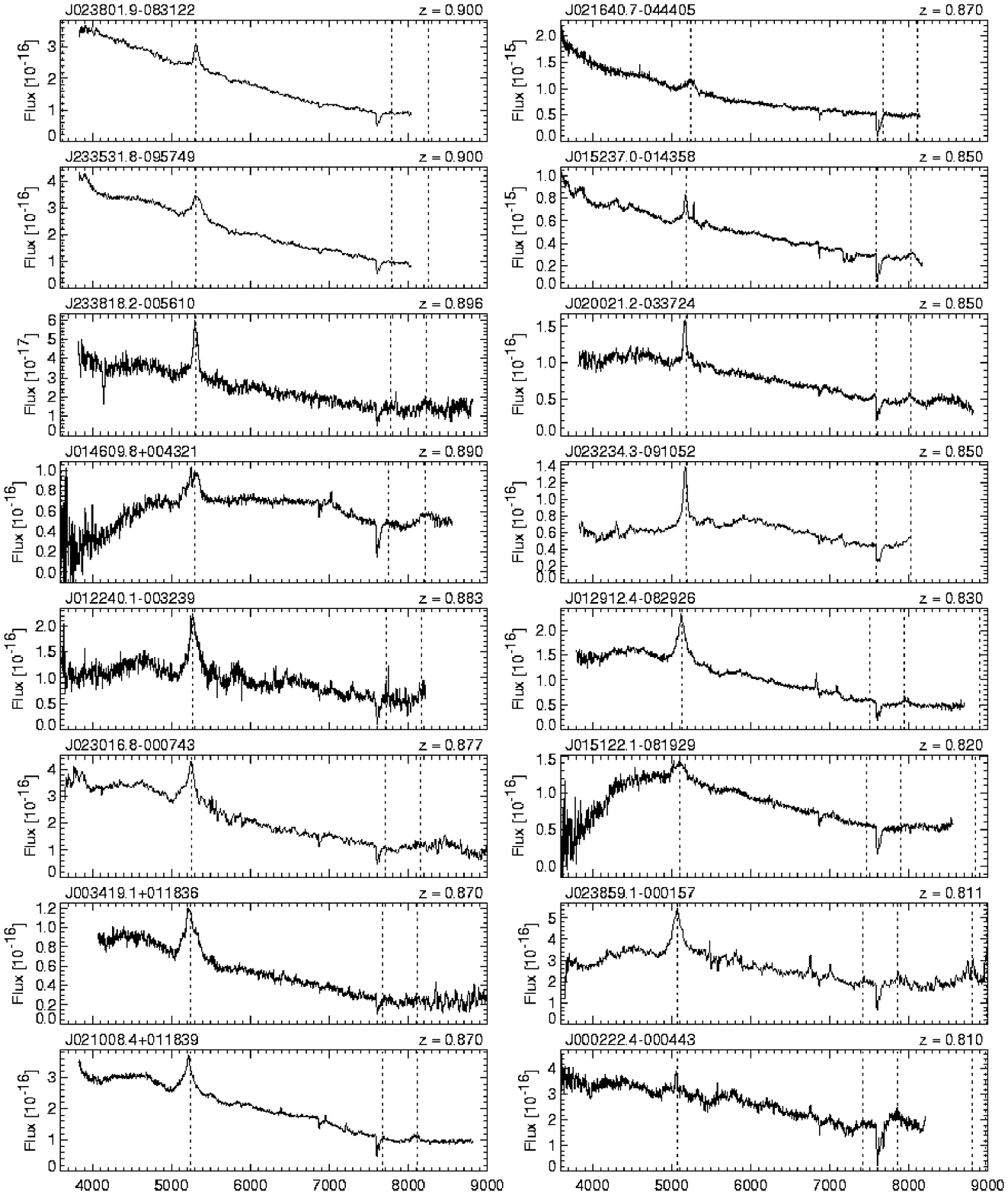}
\caption{{\it Continued.}  Spectra of FBQS quasars.}
\end{figure*}
\clearpage

\begin{figure*}[p]
\figurenum{9o}
\plotone{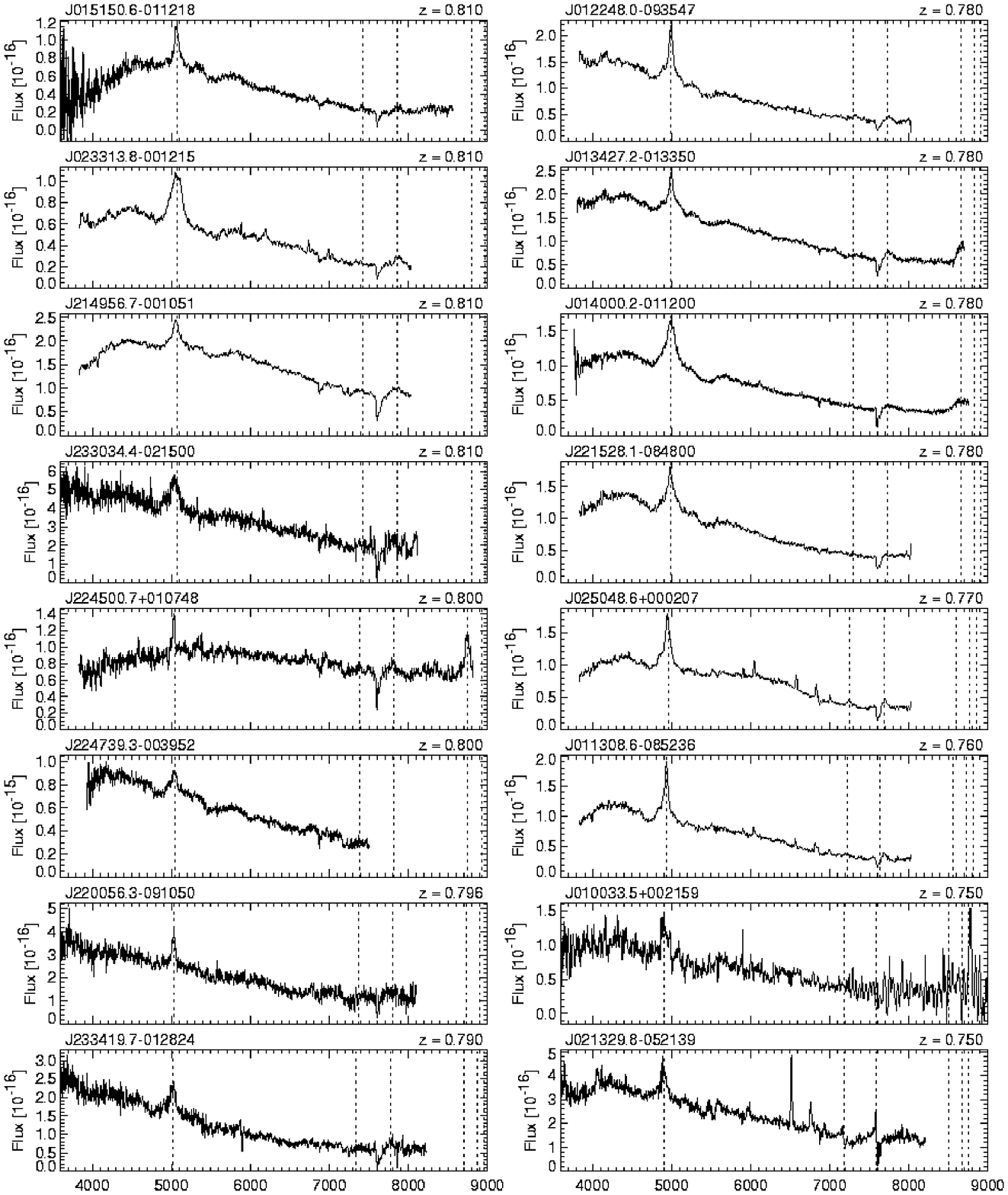}
\caption{{\it Continued.}  Spectra of FBQS quasars.}
\end{figure*}
\clearpage

\begin{figure*}[p]
\figurenum{9p}
\plotone{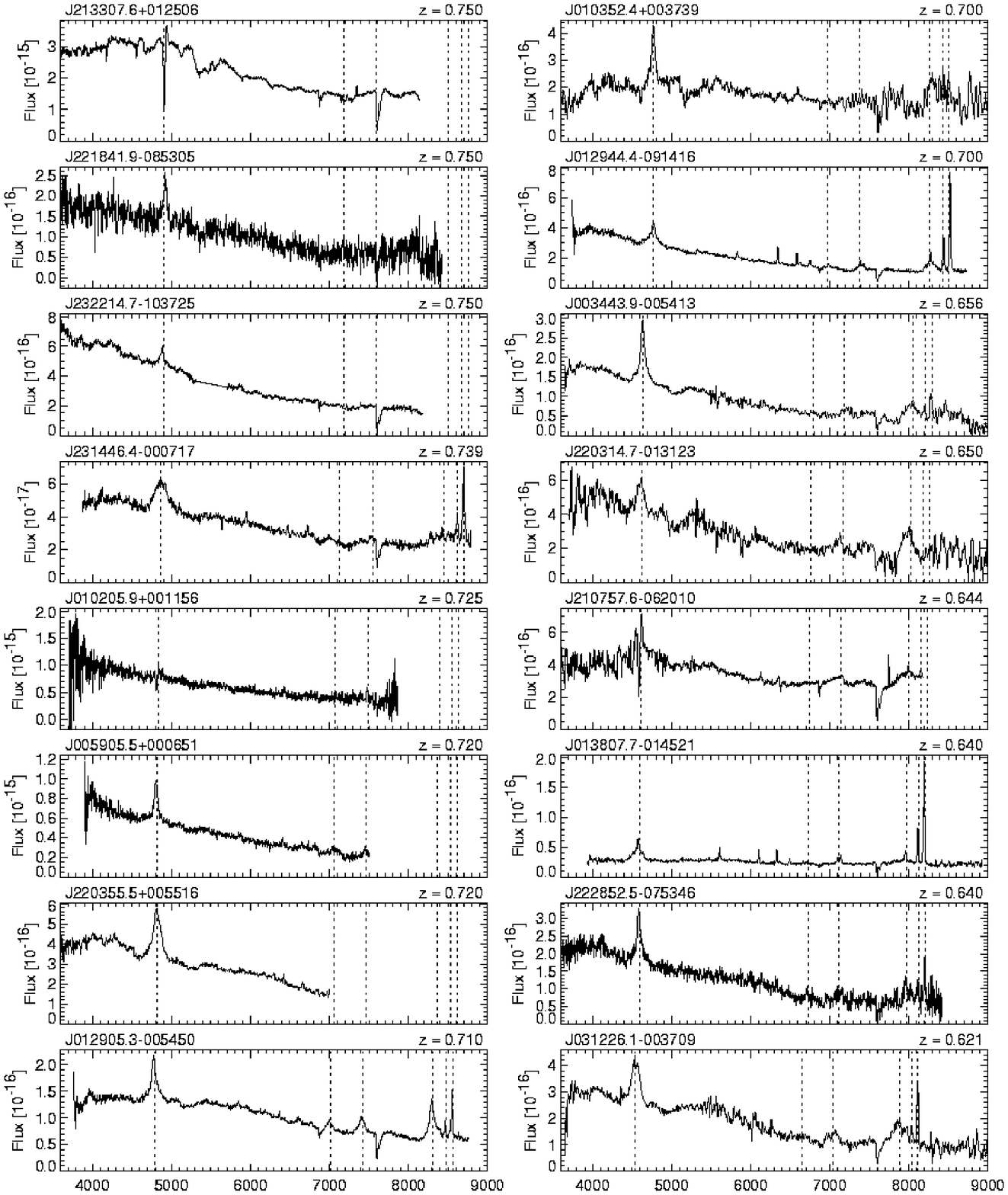}
\caption{{\it Continued.}  Spectra of FBQS quasars.}
\end{figure*}
\clearpage

\begin{figure*}[p]
\figurenum{9q}
\plotone{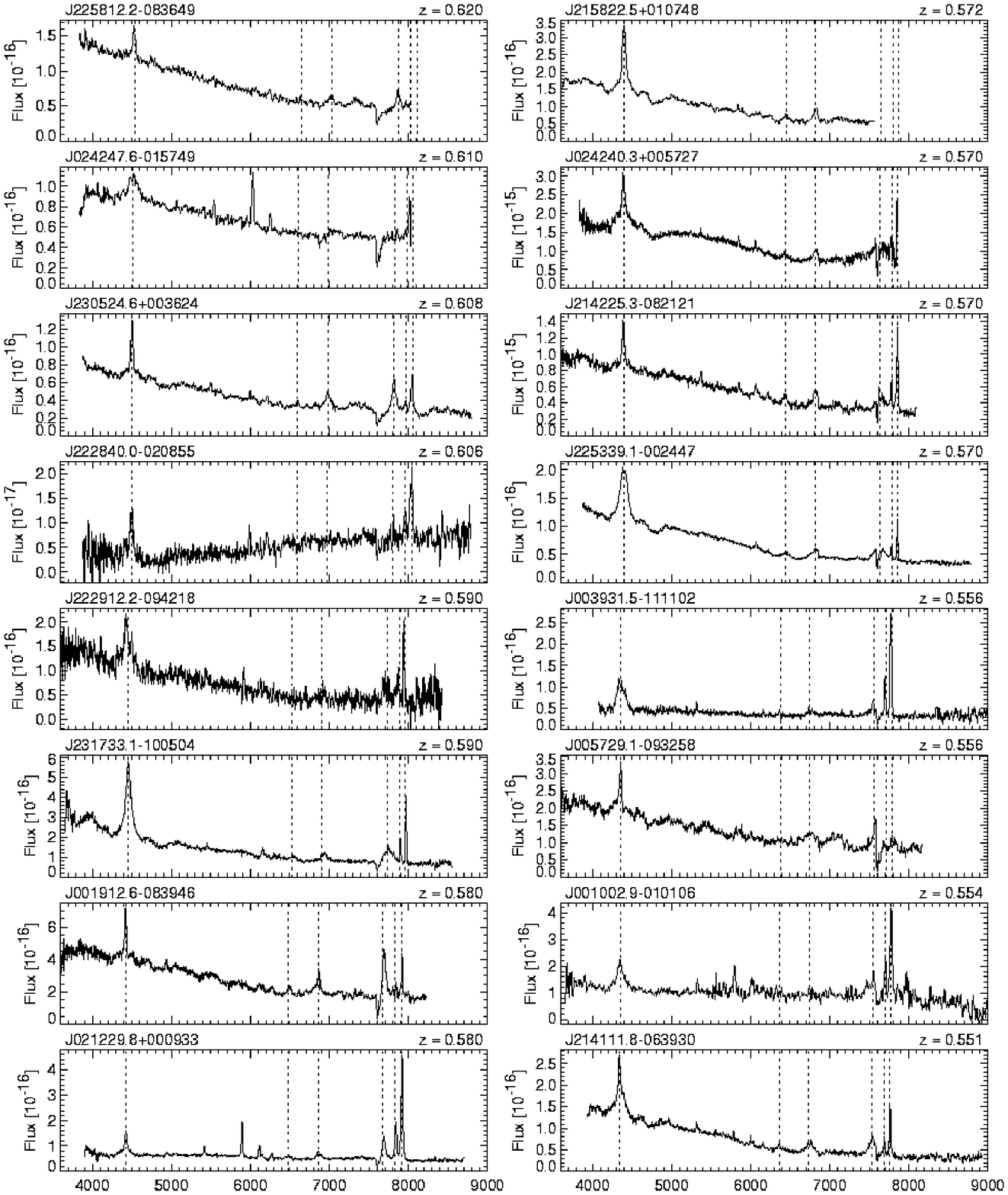}
\caption{{\it Continued.}  Spectra of FBQS quasars.}
\end{figure*}
\clearpage

\begin{figure*}[p]
\figurenum{9r}
\plotone{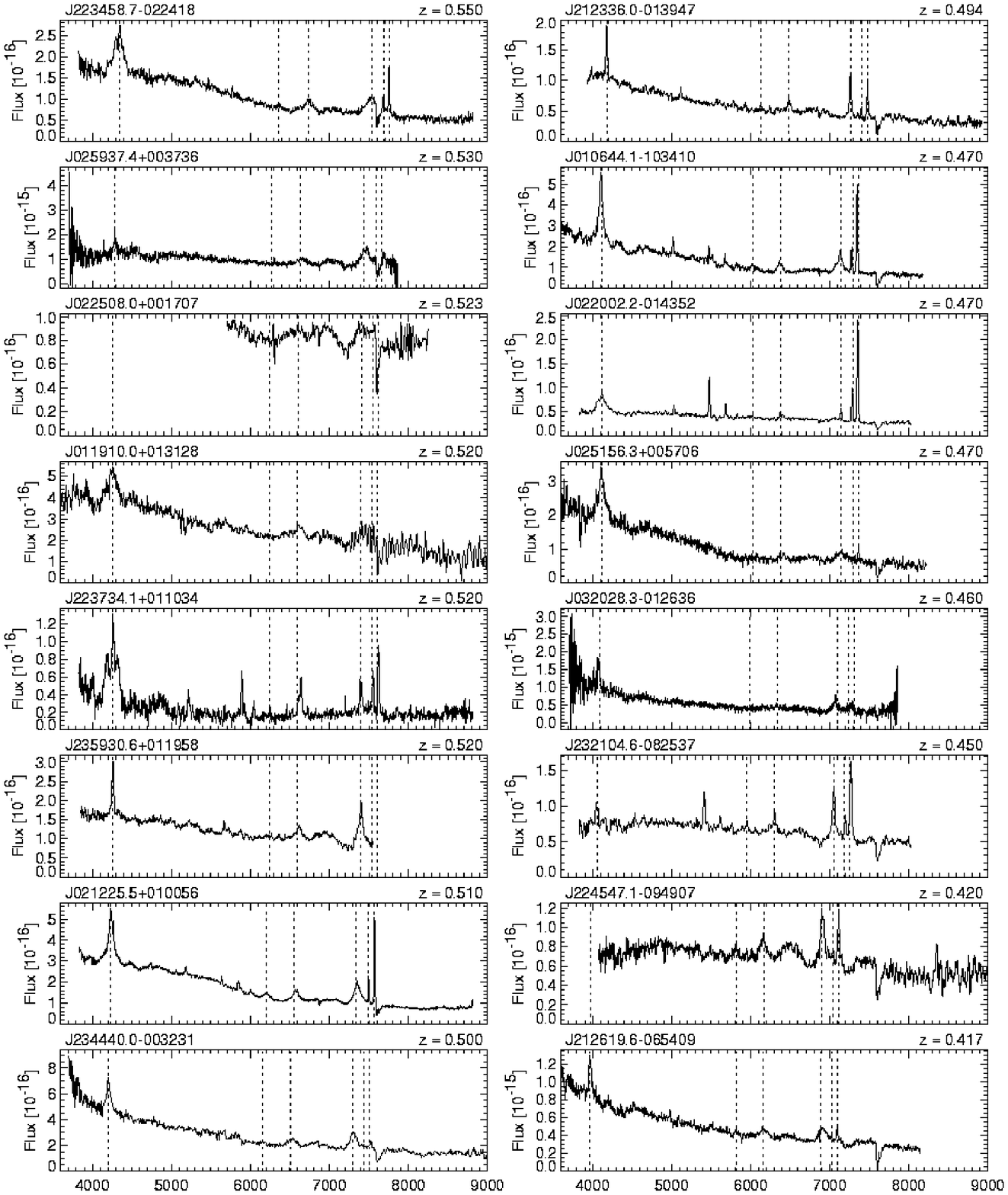}
\caption{{\it Continued.}  Spectra of FBQS quasars.}
\end{figure*}
\clearpage

\begin{figure*}[p]
\figurenum{9s}
\plotone{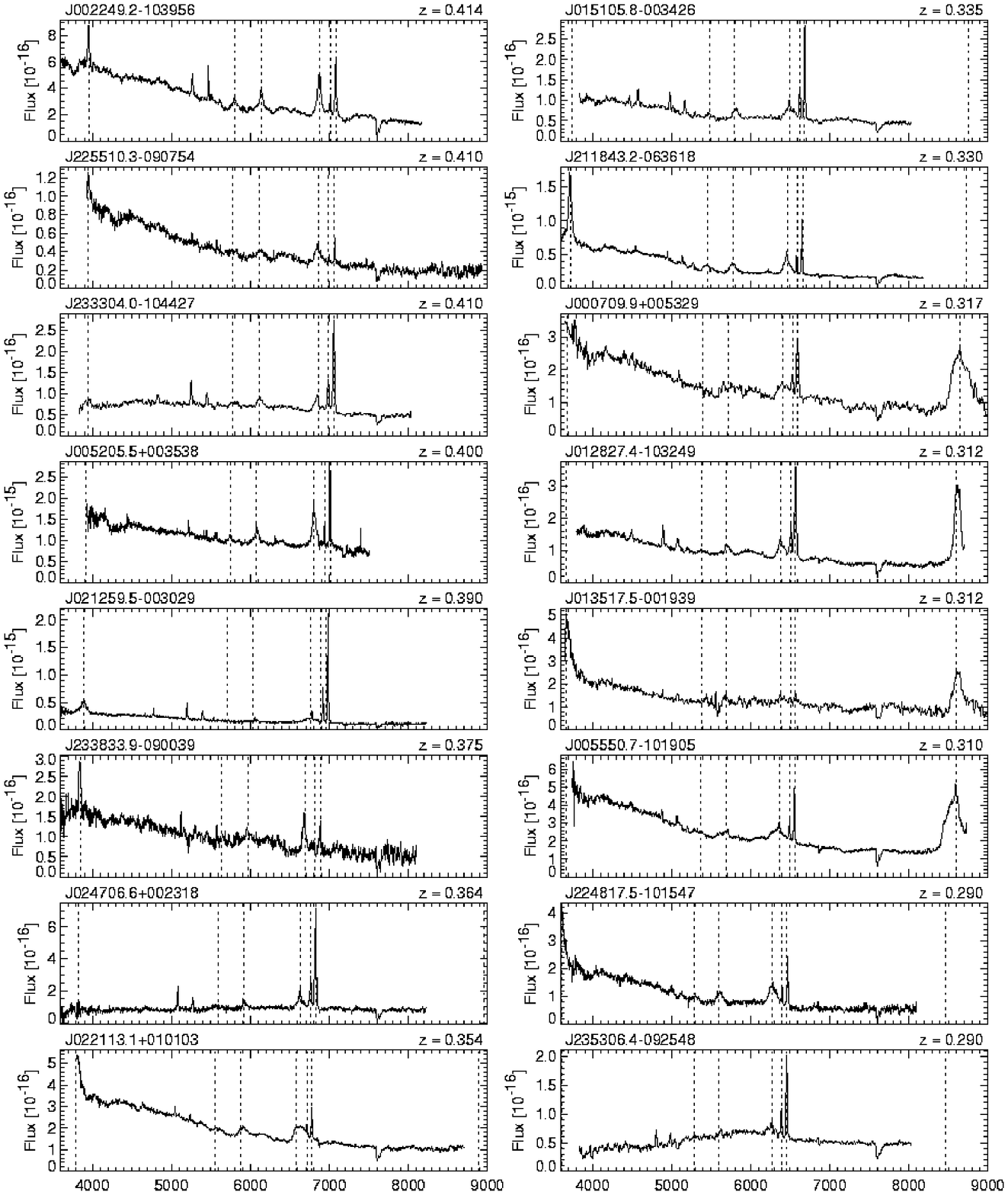}
\caption{{\it Continued.}  Spectra of FBQS quasars.}
\end{figure*}
\clearpage

\begin{figure*}[p]
\figurenum{9t}
\plotone{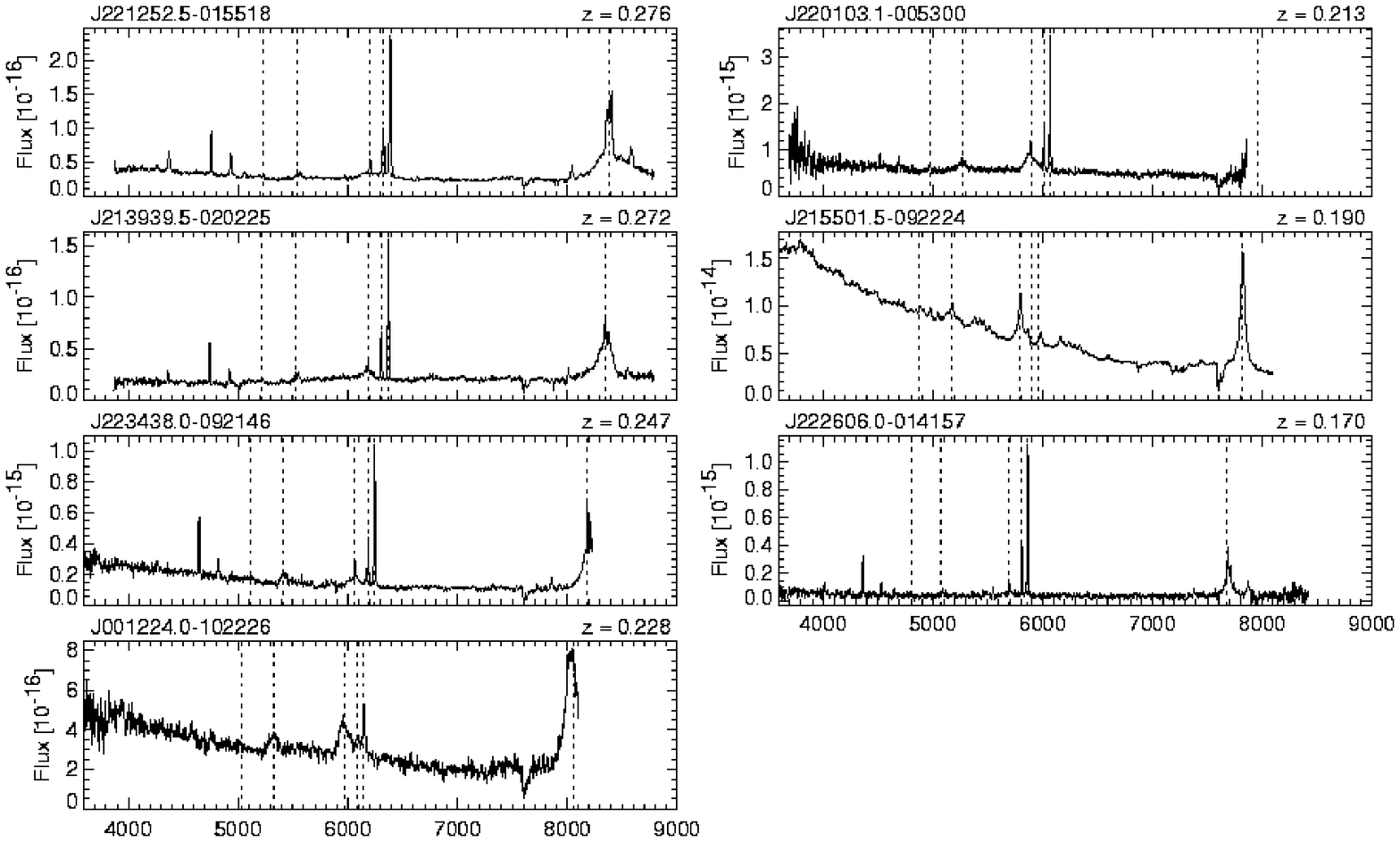}
\caption{{\it Continued.}  Spectra of FBQS quasars.}
\end{figure*}
\clearpage

\end{document}